\documentclass[lettersize,journal]{IEEEtran}
\usepackage{amsmath,amsfonts}
\usepackage{array}
\usepackage{textcomp}
\usepackage{stfloats}
\usepackage{url}
\usepackage{verbatim}
\usepackage{graphicx}
\usepackage{cite}
\usepackage{subfigure}
\usepackage{breqn}

\usepackage[T1]{fontenc}

\usepackage{algorithm}
\usepackage{algpseudocode}

\usepackage{tabularx}
\usepackage{multirow}
\usepackage{booktabs}

\usepackage[hidelinks]{hyperref}
\usepackage[dvipsnames,svgnames]{xcolor}

\usepackage{algcompatible}
\usepackage{algorithm}
\usepackage{tikz}
\usetikzlibrary{arrows,calc,intersections}
\usepackage{graphicx}
\newcommand\tikzmark[1]{%
  \tikz[overlay,remember picture,baseline] \coordinate (#1);}

\definecolor{red}{rgb}{1, 0, 0}
\definecolor{blue}{rgb}{0, 0, 1}
\newcommand\dave[1] {\textcolor{red}{\textbf{Dave: #1}}}
\newcommand\hank[1] {\textcolor{blue}{\textbf{Hank: #1}}}
\newcommand\ken[1] {\textcolor{purple}{\textbf{Ken: #1}}}
\newcommand\james[1] {\textcolor{orange}{\textbf{James: #1}}}
\newcommand\matt[1] {\textcolor{olive}{\textbf{Matt: #1}}}

\begin{document}

\definecolor{introColor}{rgb}{.8, .8, .8} 
\definecolor{begOverheadColor}{rgb}{.97, .51, .47} 
\definecolor{AdvectColor}{rgb}{.67, .88, .69} 
\definecolor{endOverheadColor}{rgb}{0.67, 0.8, 0.94} 
\definecolor{commColor}{rgb}{.98, .91, .71} 

\setlength{\fboxsep}{1pt}
\newcommand*{\Icolor}[1]{\colorbox{introColor}{#1}}
\newcommand*{\BOcolor}[1]{\colorbox{begOverheadColor}{#1}}
\newcommand*{\Acolor}[1]{\colorbox{AdvectColor}{#1}}
\newcommand*{\EOcolor}[1]{\colorbox{endOverheadColor}{#1}}
\newcommand*{\CWcolor}[1]{\colorbox{commColor}{#1}}

\newcommand*{\ModelIWord}[1]{\colorbox{introColor}{#1}}
\newcommand*{\ModelBOWord}[1]{\colorbox{begOverheadColor}{#1}}
\newcommand*{\ModelAWord}[1]{\colorbox{AdvectColor}{#1}}
\newcommand*{\ModelEOWord}[1]{\colorbox{endOverheadColor}{#1}}
\newcommand*{\ModelCWword}[1]{\colorbox{commColor}{#1}}

\def\ModelI{\Icolor{$I$}}
\def\ModelBO{\BOcolor{$BO$}}
\def\ModelA{\Acolor{$A$}}
\def\ModelEO{\EOcolor{$EO$}}
\def\ModelC{\CWcolor{$C$}}
\def\ModelW{\CWcolor{$W$}}
\def\ModelCW{\CWcolor{$CW$}}

\def\rModelI{\Icolor{$I_r$}}
\def\rModelBO{\BOcolor{$BO_{r}^{i}$}}
\def\rModelA{\Acolor{$A_{r}^{i}$}}
\def\rModelEO{\EOcolor{$EO_{r}^{i}$}}
\def\rModelCW{\CWcolor{$CW_{r}^{i}$}}

\def\rTimeModelI{\Icolor{$T_I(r)$}}
\def\rTimeModelBO{\BOcolor{$T_{BO}(r,i)$}}
\def\rTimeModelA{\Acolor{$T_{A}(r,i)$}}
\def\rTimeModelEO{\EOcolor{$T_{EO}(r,i)$}}
\def\rTimeModelC{\CWcolor{$T_{C}(r,i)$}}
\def\rTimeModelW{\CWcolor{$T_{W}(r,i)$}}
\def\rTimeModelCW{\CWcolor{$T_{CW}(r,i)$}}

\def\pModelI{\Icolor{$I_p$}}
\def\pModelBO{\BOcolor{$BO_{p}^{i}$}}
\def\pModelA{\Acolor{$A_{p}^{i}$}}
\def\pModelEO{\EOcolor{$EO_{p}^{i}$}}
\def\pModelCW{\CWcolor{$CW_{p}^{i}$}}

\def\pTimeModelI{\Icolor{$T_I(p)$}}
\def\pTimeModelBO{\BOcolor{$T_{BO}(p,i)$}}
\def\pTimeModelA{\Acolor{$T_{A}(p,i)$}}
\def\pTimeModelEO{\EOcolor{$T_{EO}(p,i)$}}
\def\pTimeModelC{\CWcolor{$T_{C}(p,i)$}}
\def\pTimeModelW{\CWcolor{$T_{W}(p,i)$}}
\def\pTimeModelCW{\CWcolor{$T_{CW}(p,i)$}}


\title{Parallelize Over Data Particle Advection: Participation, Ping Pong Particles, and Overhead}

\author{
    \IEEEauthorblockN{Zhe Wang\IEEEauthorrefmark{1}, Kenneth Moreland\IEEEauthorrefmark{1}, Matthew Larsen\IEEEauthorrefmark{2}, James Kress\IEEEauthorrefmark{3}, Hank Childs\IEEEauthorrefmark{4},
     David Pugmire\IEEEauthorrefmark{1}}
    \\\IEEEauthorblockA{\IEEEauthorrefmark{1}Oak Ridge National Laboratory, Oak Ridge, TN, USA \\\ }
    \IEEEauthorblockA{\IEEEauthorrefmark{2}Luminary Cloud, Inc., CA, USA \\\ }
    \IEEEauthorblockA{\IEEEauthorrefmark{3}King Abdullah University of Science \& Technology (KAUST), Saudi Arabia \\\ }
    \IEEEauthorblockA{\IEEEauthorrefmark{4}University of Oregon, Eugene, OR, USA \\\ }
\vspace{-30pt}
\thanks{This manuscript has been authored by UT-Battelle, LLC under Contract No. DE-AC05-00OR22725 with the U.S. Department of Energy. The publisher, by accepting the article for publication, acknowledges that the U.S. Government retains a non-exclusive, paid up, irrevocable, world-wide license to publish or reproduce the published form of the manuscript, or allow others to do so, for U.S. Government purposes. 
The DOE will provide public access to these results in accordance with the DOE Public Access Plan 
(http://energy.gov/downloads/doe-public-access-plan).
}
}

\maketitle
%




\begin{abstract}

Particle advection is one of the foundational algorithms for visualization and analysis and is central to understanding vector fields common to scientific simulations.
Achieving efficient performance with large data in a distributed memory setting is notoriously difficult.
Because of its simplicity and minimized movement of large vector field data, the \emph{Parallelize over Data} (POD) algorithm has become a de facto standard.
Despite its simplicity and ubiquitous usage, the scaling issues with the POD algorithm are known and have been described throughout the literature.
In this paper, we describe a set of in-depth analyses of the POD algorithm that shed new light on the underlying causes for the poor performance of this algorithm.
We designed a series of representative workloads to study the performance of the POD algorithm and executed them on a supercomputer while collecting timing and statistical data for analysis.
we then performed two different types of analysis. In the first analysis, we introduce two novel metrics for measuring algorithmic efficiency over the course of a workload run. The second analysis was from the perspective of the particles being advected.
%
Using particle-centric analysis, we identify that the overheads associated with particle movement between processes (not the communication itself) have a dramatic impact on the overall execution time. These overheads become particularly costly when flow features span multiple blocks, resulting in repeated particle circulation (which we term ``ping pong particles'') between blocks. 
Our findings shed important light on the underlying causes of poor performance and offer directions for future research to address these limitations.

\end{abstract}  


\section{Introduction}

Particle advection is a foundational algorithm for the visualization and analysis of flow fields.
Particle advection in fluid flow visualization starts with a vector field representing the velocity of a fluid and traces the trajectories that massless particles would take if carried by this fluid.
These trajectories can be used directly to represent the flow as streamlines, but they also form the basis of numerous other flow visualization algorithms, including stream surfaces \cite{Hultquist1992,Wijk1993}, Lagrangian coherent surfaces \cite{Guo2016}, and Poincar\'{e} plots \cite{Loeffelmann1998,Sanderson2010}.

Tracing particles can be computationally expensive, and although the trajectories of particles are independent of each other, efficient parallel processing of trajectory computations can be difficult.
The most common approach to parallel particle advection over distributed memory ranks (such as in an MPI job \cite{MPI}) is \emph{Parallelize over Data} (POD), which assigns a partition of the flow field (a ``block'') to each compute rank and pushes trajectory computation to the rank containing the particle's position.
POD is popular because it simplifies the data distribution and execution of the algorithm, and it also minimizes data movement, which is particularly crucial for \emph{in situ} visualization \cite{Sane2021}.
POD also matches the data distribution model commonly used by data parallel software frameworks for scientific visualization \cite{Moreland2013:TVCG}. Therefore, POD is the default algorithm used by scalable visualization tools such as ParaView \cite{ParaView}, VisIt \cite{VisIt}, and Ascent~\cite{larsen2022ascent}.

Although straightforward to implement and sometimes efficient, POD can become slow when algorithm configurations or data sets lead to unbalanced particle advection workloads.
The fundamental problem with POD is that the division of work is dictated by the structure of the velocity field rather than the amount of computation to be done, and thus work becomes unbalanced when particles distribute themselves unevenly to velocity field blocks.
This can occur for a variety of reasons, including the following.
First, seed placement in a small region of the overall volume will limit the number of ranks that can participate in performing work. Additional ranks will only get work as the seeds propagate through the volume. This is of course dependent on the nature of the velocity field and some ranks may see little or no work.
Second, if the velocity field contains one or more sinks, the particles will be attracted to these data blocks, and hence ranks that contain these features.
Even if the initial seed locations are distributed throughout the entire volume, the particles, and hence the work to advect them, become concentrated by the ranks with sink features that collect them.

%

In this paper, we present new results that shed important light on the underlying causes of imbalance and decreased performance for distributed memory particle advection.
We design a set of experiments to explore the performance of this algorithm across a diverse set of workloads.
From the different computational tasks within the algorithm, we formulate a model that describes the execution time and is used as the basis for our analysis of results.
Further, we instrument the code to collect a variety of timings, counters, and statistics from our model to quantify the algorithmic behavior across these differing workloads.

The contributions of this paper are described below.
We introduce a novel metric called \emph{rank participation} to effectively quantify the evolving workload imbalances of particle advection over time. This metric serves as an efficient tool for measuring the degree of associated imbalance between different workloads on the same dataset as well as among different datasets. We additionally introduce the concept of \emph{aggregated rank participation} to facilitate the comparison of workload imbalances across diverse datasets and their associated settings.
The challenges of the POD algorithm at scale are well known~\cite{EuroVisSTAR2023}.
In using these metrics, we find that the POD algorithm can be efficient for some select cases, but that the participation rates are much worse than expected for the majority of the configurations in our experiments.

Our second contribution comes from an analysis of individual particles being advected. We have identified two primary factors that lead to decreased performance, which to the best of our understanding have not been described in the literature.
First, the number of particles that are processed together in a group on a given rank can cause a significant increase in the overall execution time.
Second, the overhead (not the communication itself) of transferring particles between processes grows unexpectedly and has a dramatic impact on the overall execution time. We also identify cases where these overheads rapidly accumulate. These cases, which we term ``ping pong effect'', occur when flow features span multiple blocks. This results in particles repeatedly circulating between the blocks containing the flow features.

The rest of the paper is organized as follows. \S~\ref{sec:related-work} discusses the background and related work. \S~\ref{sec:model} describes the particle advection algorithm and derives a model for the execution time.
\S~\ref{sec:exp_overview} describes the experiments performed, and  \S~\ref{sec:expresults} provides a detailed evaluation of the results.
In \S~\ref{sec:discussion}, we discuss the implications of our findings and describe approaches for alleviating their effects.
Finally, we conclude the paper and indicate future works in \S~\ref{sec:con_future}.

\if
Although straightforward to implement and sometimes efficient, the fundamental problem with POD is that the division of work is dictated by the structure \fix{(decomposition?)} of the velocity field's mesh 
rather than the amount of computation to be done.
If a significant fraction of particles move to a sink or small vortex in the fluid, the rank containing that feature will be overloaded while the remaining ranks will remain idle.
This idle time \fix({imbalance?}) in turn jeopardizes the ability for POD to run efficiently.
The goal of this \fix{short} paper is to better understand the rate at
which imbalances occur over a variety of scenarios.

In this work we explore the \emph{rank participation} of POD particle advection.
The rank participation is the fraction of ranks actively doing work.
If $N_T$ is the total number of ranks in the parallel job and $N_A$ is the number of ranks actively doing work, then the rank participation is $N_A/N_T$.
The rank participation can be measured instantaneously or integrated over a run.
A rank participation of $1$ represents a perfectly load balanced operation whereas a small rank participation of $1/N_T$ means one rank is working while the rest remain idle.
The two charts on the right hand side of Figure~\ref{fig:gant_rp} show the rank participation over time for two of the runs in our experiment.
\fi

\section{Background and Related Work}
\label{sec:related-work}

\subsection{Parallel particle advection algorithms}
When advecting particles in a distributed-memory parallel system (i.e., one using a Message Passing Interface (MPI) \cite{MPI} model), there are two basic approaches to the implementation~\cite{Pugmire:HPV}.
The first is a data parallel \emph{Parallelize over Data} (POD) approach where data are partitioned and distributed among process ranks to manage particles in those domains.
The second is a task parallel \emph{Parallelize over Seeds} (POS) approach where particles are evenly distributed among process ranks and data are loaded on ranks as needed.
These approaches can also be combined to form hybrid implementations \cite{Pugmire2009,kendall2011simplified,Peterka2011:IPDPS,guo2013coupled, binyahib2019lifeline,binyahib2021hylipod}.
We direct the reader to Zhang and Yuan \cite{Zhang2018} for an overview and Yenpure~\cite{yenpure2023state} for optimization strategies of parallel particle tracing systems.

Recent computer architecture introduces computer processors such as multi-core CPUs and GPUs that have a high degree of parallelism.
Particle advection systems taking advantage of these processors have shared memory access, and thus, a POS approach is a natural fit for local parallelism \cite{Pugmire2018}.
That said, in a high-performance computer, multiple parallel processors are joined in a cluster configuration, giving rise to what is often known as an MPI+X configuration.
Particle advection in these systems uses a POS approach locally but a basic POD approach across the process ranks of the distributed system \cite{Camp2010,Camp2013}.
In such a configuration, it is common to perform both the advection and communication of particles in batches. That is, advecting all of the particles that are present on a particular node followed by communication of particles to the next destination.


\subsection{Performance evaluation of particle advection algorithms}

Peterka et al.~\cite{Peterka2011:IPDPS} utilize a particle tracer to monitor key metrics of the parallel particle advection system. They discuss several bottlenecks of the particle advection system.
Our work views the execution time of the whole particle advection system from both perspectives of computation ranks and long-running particles, which better explains the bottleneck of the particle advection system. We also derive metrics that evaluate the degrees of corresponding factors causing the bottleneck.

Childs et al. \cite{Childs2014} explore the relationship between particle advection workloads and the high-performance hardware in which they are run. They conclude that powerful processors and GPUs can help with dense particle workloads, but those benefits diminish as the workload lightens. Our study uses the performance model as an intermediate tool to explore the reason causing workload imbalance. 

Sisneros et al.~\cite{sisneros2016tuned} vary multiple parameters, and evaluate the associated performance of parallel particle advection based on POD. Their results reveal that the default configurations used in production-ready visualization tools are not always optimal. However, their analysis only shows the relationship between the configuration and performance, and they do not discuss the underlying reason causing poor performance for a specific configuration. Our work uses metrics associated with each process and long-running particles to reveal how the unbalanced workload decreases the performance of parallel particle advection.

Binyahib et al.~\cite{binyahib2020parallel} present a comprehensive evaluation to show how different particle advection algorithms behave with different workload factors. Their results show that POD has the best performance when using a large seeding box. Our work further explores situations that cause a poor performance of the POD algorithm with a large seeding box.

\subsection{Load-balanced particle advection algorithms}
Nouanesengsy et al.~\cite{nouanesengsy2011load} describe a partitioning algorithm based on workload estimation to reduce the idle time of each rank for a parallel particle streamline generation algorithm. Subsequent research \cite{zhang2018dynamic,xu2022reinforcement} use dynamic load balancing strategies to adjust the workloads between processes to improve the performance of particle tracing systems. By analyzing the reasons that cause that imbalance of particle advection in detail, our work complements the literature that studies the workload participation strategy of parallel particle advection. In particular, the rank participation value shows how load balancing between different processes changes during the particle advection, and the long-running particle statistics show how ``ping pong particles'' can cause workload imbalance. In addition, we focus on asynchronous MPI communication, which is barely discussed in the aforementioned related works for load-balanced particle advection systems. 










\section{Theoretical Considerations}
\label{sec:model}
This section presents a framework used to analyze the execution time for a parallel particle advection workload. In particular,
\S~\ref{subsec:model:algorithm} describes the POD algorithm adopted in this study. \S~\ref{subsec:model:def} defines terms for computational tasks in the POD algorithm and provides a model for the execution time from a rank- and a particle-based perspective. The model provides a theoretical foundation in explaining experiment results in \S~\ref{sec:expresults}.

\subsection{Particle advection algorithm}
\label{subsec:model:algorithm}



\begin{algorithm}[ht]
\caption{Parallel particle advection algorithm.}\label{pvalgo}

\begin{algorithmic}[1]
\vspace{5pt}
\State\tikzmark{intro0}\hspace{-1pt}\colorbox{introColor}{\makebox[2.8in][l]{\textbf{/* Initialization */}}}
\State ActiveQueue = GetSeeds()\label{pvalgo:I1}
\State Count = TotalNumberOfParticles()\tikzmark{intro1}\label{pvalgo:I2}
\vspace{5pt}

\WHILE{Count $> 0$}\label{pvalgo:whileBegin}
    \State $P_{out}$ = Empty
    \State $N_{term}$ = 0
    \IF{ActiveQueue not Empty}
        \vspace{2pt}
        \State\tikzmark{BO0}\hspace{-1pt}\colorbox{begOverheadColor}{\makebox[2.5in][l]{\textbf{/* Begin overhead */}}}
        \State Initialize Kernel with ActiveQueue\tikzmark{BO1}\label{pvalgo:BO1}
\vspace{5pt}
        
        \State\tikzmark{Advect0}\hspace{-1pt}\colorbox{AdvectColor}{\makebox[2.5in][l]{\textbf{/* Advection */}}}
        \State Result = Kernel execute\tikzmark{Advect1}\label{pvalgo:A1}
\vspace{5pt}
        \State\tikzmark{EO0}\hspace{-1pt}\colorbox{endOverheadColor}{\makebox[2.5in][l]{\textbf{/* End Overhead */}}}
        \State $P_{out}$ = Exiting particles from Result\label{pvalgo:EO1}
        \State $N_{term}$ = Num terminated from Result\label{pvalgo:EO2}
        \State ActiveQueue = Empty\tikzmark{EO1}\label{pvalgo:EO3}
    \ENDIF
\vspace{2pt}
    \State\tikzmark{C0}\hspace{-1pt}\colorbox{commColor}{\makebox[2.65in][l]{\textbf{/* Communication and Wait */}}}   
    \State Send $P_{out}$\label{pvalgo:C1}
    \State Count = UpdateCounter($N_{term})$\label{pvalgo:C2}
    \State $P_{in}$ = Receive incoming particles\label{pvalgo:C3}
    \State Add $P_{in}$ to ActiveQueue\tikzmark{C1}\label{pvalgo:C4}
    \vspace{3pt}

\ENDWHILE\label{pvalgo:whileEnd}

\end{algorithmic}
\end{algorithm}

 \begin{tikzpicture}[remember picture,overlay]
    \draw[line width=1pt, draw=introColor, rounded corners=0pt, fill=introColor, fill opacity=0.05]
        ([xshift=-1pt,yshift=8pt]intro0.north) rectangle ([xshift=62pt,yshift=-3pt]intro1.south);

    \draw[line width=1pt, draw=begOverheadColor, rounded corners=0pt, fill=begOverheadColor, fill opacity=0.05]
        ([xshift=-1pt,yshift=8pt]BO0.north) rectangle ([xshift=39pt,yshift=-3pt]BO1.south);   

    \draw[line width=1pt, draw=AdvectColor, rounded corners=0pt, fill=AdvectColor, fill opacity=0.05]
        ([xshift=-1pt,yshift=8pt]Advect0.north) rectangle ([xshift=83pt,yshift=-3pt]Advect1.south);    

    \draw[line width=1pt, draw=endOverheadColor, rounded corners=0pt, fill=endOverheadColor, fill opacity=0.05]
        ([xshift=-1pt,yshift=8pt]EO0.north) rectangle ([xshift=90pt,yshift=-3pt]EO1.south);

    \draw[line width=1pt, draw=commColor, rounded corners=0pt, fill=commColor, fill opacity=0.05]
        ([xshift=-1pt,yshift=8pt]C0.north) rectangle ([xshift=90pt,yshift=-3pt]C1.south);

\end{tikzpicture}

Algorithm~\ref{pvalgo} contains the pseudocode for performing parallel particle advection used in this paper.
It uses the parallelize over data (POD) scheme where the data blocks are spatially partitioned across a set of processes. Each rank performs the advection for each particle that passes through the block it is assigned. 
During \Icolor{initialization}, each rank assigns the particles contained in the local block to a variable called ActiveQueue (Line~\ref{pvalgo:I1}) and then the number of particles across all processes is computed (Line~\ref{pvalgo:I2}).

Following initialization, the algorithm executes a while loop until all of the particles terminate or exit the global bounds of the data. The while loop consists of two phases: computation and communication.
%
%

The computation phase is performed if the rank has any work.
If the ActiveQueue is not empty, a particle advection kernel is \BOcolor{initialized} with the particles in ActiveQueue (Line~\ref{pvalgo:BO1}).
Execution of the kernel will advance each particle by solving the differential equation using a Runge-Kutta 4 (RK4) solver~\cite{NumericalRecipesInC} to compute the path (Line~\ref{pvalgo:A1}).
Each particle is \Acolor{advected} until a termination criterion is met (i.e., the maximum number of iterations, or the particle enters a zero-velocity region-- a sink), or it exits the spatial domain of the data block.
Next, at the \EOcolor{end} of advection, the particles exiting the block, $P_{out}$, and the number of terminations are extracted from the advection Result (Lines~\ref{pvalgo:EO1},~\ref{pvalgo:EO2}) and the ActiveQueue is cleared (Line~\ref{pvalgo:EO3}).

Following computation, the \CWcolor{communication} phase is performed using asynchronous message passing with MPI~\cite{mpi40}.
The first step is to send particles in the $P_{out}$ queue to their destinations. The destination rank can be determined from the location of each particle in $P_{out}$.\footnote{The states of particles are also updated when computing terminated particles in Line~\ref{pvalgo:EO2}.} The particles being sent to each destination are bundled together and sent using the asynchronous \texttt{MPI\_Isend}  call (Line~\ref{pvalgo:C1}).
The second step is to update the global counter for the locally terminated particles (Line~\ref{pvalgo:C2}).
If $P_{out}$ is empty or $N_{term}$ is zero, then the operations are both no-ops.
The third step (Line~\ref{pvalgo:C3}) is to receive any incoming particles that have been sent by other ranks. This is done using \texttt{MPI\_Iprobe}, which will indicate if any messages have arrived from other ranks.
If any messages have asynchronously arrived, the \texttt{MPI\_Irecv} call is used to receive them. These incoming particles are added to the ActiveQueue (Line~\ref{pvalgo:C4}) for processing during the next iteration of the loop.

\subsection{Execution time model}
\label{subsec:model:def}

From Algorithm\ref{pvalgo} we define the following terms.
\begin{itemize}
    \item \ModelI\  is the time for initialization. (Lines~\ref{pvalgo:I1}-~\ref{pvalgo:I2})
    \item \ModelBO\   denotes ``begin overhead'', the overhead required to initialize a worklet with a batch of particles. This overhead includes the time to select active particles and create the necessary data structures within the kernel for particle advection. (Line~\ref{pvalgo:BO1}).
    \item \ModelA\ denotes the time to advect a batch of particles until they terminate or leave the data block (Line~\ref{pvalgo:A1}).
    \item \ModelEO\  denotes ``end overhead'', the overhead required to determine the status of each particle%
    , organize the data into outgoing buffers, and clearing the ActiveQueue (Lines~\ref{pvalgo:EO1}-~\ref{pvalgo:EO3}).
    
    \item \ModelC\ represents the time to exchange particles and update the global counter. This includes serializing the particles into a message, sending it over the network, and deserialization. (Lines~\ref{pvalgo:C1}-~\ref{pvalgo:C3}).
    \item \ModelW\ represents the time spent when a rank has no work to perform and waiting for particles to arrive. When using asynchronous communication this term is difficult to compute, so we merge \ModelC\ and \ModelW\ into a single term, \ModelCW\ that captures the communication and wait time.    
\end{itemize}

From the perspective of a particular rank, we can model the execution time as follows.
Using the terminology in previous works~\cite{Peterka2011:IPDPS, Pugmire:HPV}, each iteration of the while loop (Lines~\ref{pvalgo:whileBegin}-~\ref{pvalgo:whileEnd} in Algorithm~\ref{pvalgo}) is called a \emph{round}.
For rank $r$, and a total number of rounds (iterations) $N_r$, the execution time, $T_r$ can be modeled as follows. 
\begin{dmath}
T_r = \rModelI\ + \sum\limits_{i=1}^{N_r} \left [ \rModelBO + \rModelA + \rModelEO  + \rModelCW \right ]
\label{model:eq1}
\end{dmath}

The total execution time for the entire
workload with $R$ ranks, $T_{total}$ is the 
execution time for the rank with the longest execution time.
\begin{equation}
   T_{total} = \max (T_1,  \ldots, T_{R}) \nonumber
\end{equation}

As a particle is advected within a dataset, it traverses through a sequence of blocks. Given a particle with index $p$ that travels through a total of $N_p$ blocks (i.e., $B_1$, \ldots , $B_{N_p}$). For the $p^{th}$ particle that traverses through a total of $N_p$ blocks, the execution time $T_p$ can be modeled as follows.

\begin{dmath}
T_p = \pModelI + 
\sum\limits_{i=1}^{N_p} \left [ \pModelBO +
\pModelA + 
\pModelEO + 
\pModelCW
\right ]
\label{model:eq2}
\end{dmath}

The total execution time for the entire workload of $P$ particles, $T_{total}$ is simply the execution time for the slowest particle.
\begin{equation}
   T_{total} = \max (T_1, \ldots, T_P) \nonumber
\end{equation}

It is worth noting that Algorithm~\ref{pvalgo} processes groups of particles together. This design is desirable to keep processors busy and to minimize the amount of communication. As such, particles are processed in groups, and each term included in $T_p$ in Equation~\ref{model:eq2} is influenced by all particles in the same group as particle $p$.

  
\section{Experimental Overview}
\label{sec:exp_overview}

Our experiments are designed to explore the magnitude and associated reasons of imbalance for the POD particle advection algorithm using a diverse set of workloads. \S~\ref{subsec:model:metrics} describes the metrics adopted in our evaluation, and \S~\ref{subsec:exp_setup} describes the experimental setup, including the datasets, workloads, and evaluation platforms. \S~\ref{subsec:exp_data_collection} details how associated metrics used for analysis data are collected in experiments.

\subsection{Evaluation goals and metrics}
\label{subsec:model:metrics}
To represent the imbalance of asynchronous particle advection and show the reason that causes the imbalance, we derive multiple metrics based on the theoretical considerations discussed in \S~\ref{sec:model}. Specifically, metrics like \emph{weak scaling efficiency} and \emph{rank participation} are used to explain the relationships between poor scalability and workload imbalance. Meanwhile, the \emph{aggregated participation rate} metric facilitates comparisons of workload imbalances across diverse datasets and configurations. Furthermore, \emph{particle statistics} can explain the attributes of long-running particles and what reasons cause a longer execution time than other particles. Detailed descriptions of these metrics are listed as follows.

\emph{\textbf{Weak scaling efficiency}} provides an overall view for a workload's scalability. Low weak scaling efficiency indicates that the workload becomes more unbalanced with an increase in the number of processors during the particle advection.

\emph{\textbf{Rank participation}} represents the ratio of busy ranks involved in the particle advection.
For any moment $t$ and the specific rank $r$, if rank $r$ is in the advection stage (namely, \ModelBO\, \ModelA\ or \ModelEO\ in Equation~\ref{model:eq1}), the rank participation value for this rank  ($RP_r$) equals $1$; otherwise, it is zero. The rank participation value for all ranks is $\sum_{r=1}^{N}RP_r/N$, where $N$ is the total number of ranks.
Given that rank participation is measured at an instantaneous time $t$, it can support the case where asynchronous POD has unaligned rounds.

\emph{\textbf{Aggregated participation rate}} represents the overall status of the particle participation for a complete particle advection run and is measured as the average rank participation over the run.
If rank participation is represented as a continuous function over the time of the run, then the aggregated participation rate is the area under this curve divided by the time of the run.
In practice, the rank participation is measured at discrete intervals, and these measurements are averaged.

\emph{\textbf{Particle statistics}} include a series of information associated with particles shown in Equation~\ref{model:eq2}: particle alive time, number of traversed compute ranks, particle termination reason, particle accumulated advection time, overhead, and wait time. By analyzing these statistical metrics we can understand the wide variation in run time among particles.
Associated analysis results can identify the root reason for workload imbalance and guide us in choosing appropriate solutions to improve the performance of particle advection further.







\subsection{Experimental setup}
\label{subsec:exp_setup}

\begin{figure*}[ht]
\centering
\subfigure[Tokamak]{
\label{fig:fusion}  \includegraphics[width=30mm]{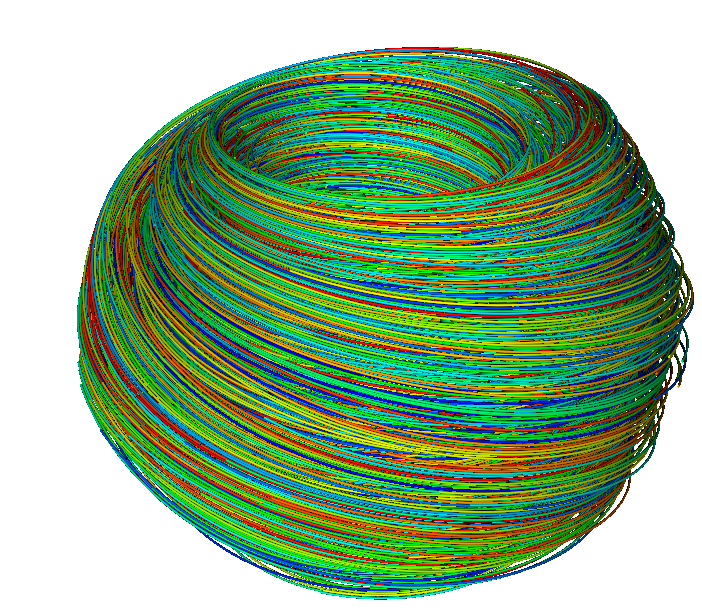}}
\subfigure[Supernova]{\label{fig:astro}  \includegraphics[width=30mm]{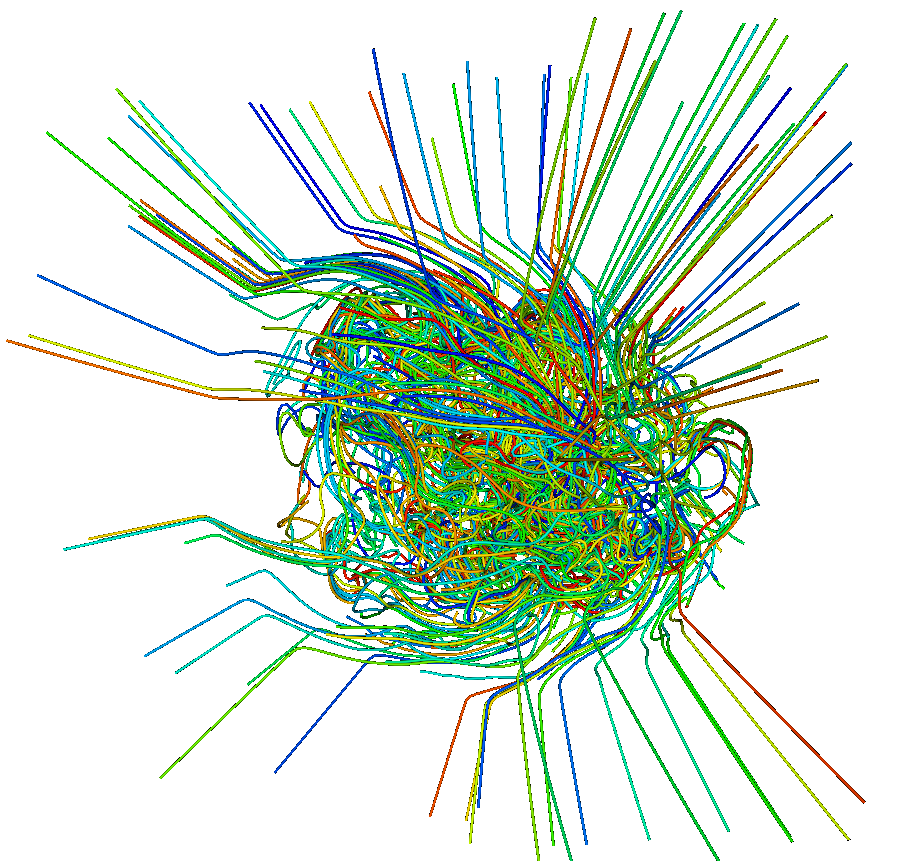}}
\subfigure[Hydraulics]{\label{fig:fishtank}  \includegraphics[width=30mm]{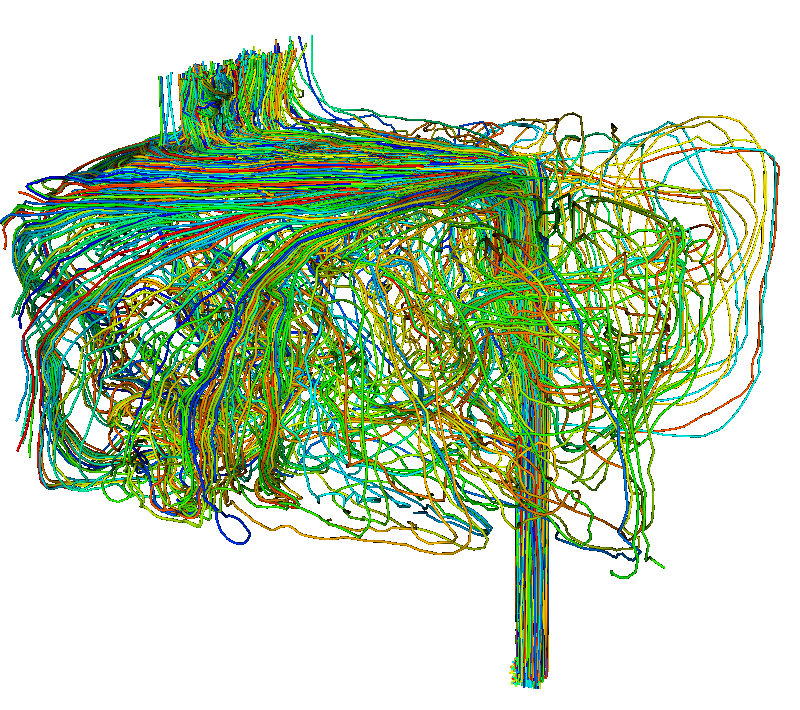}}
\subfigure[CloverLeaf3D]{\label{fig:cloverleaf}  \includegraphics[width=30mm]{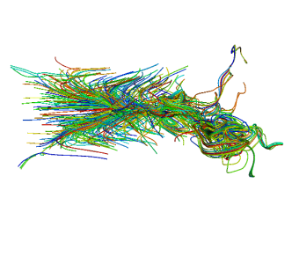}}
\subfigure[Synthetic]{\label{fig:syn}  \includegraphics[width=30mm]{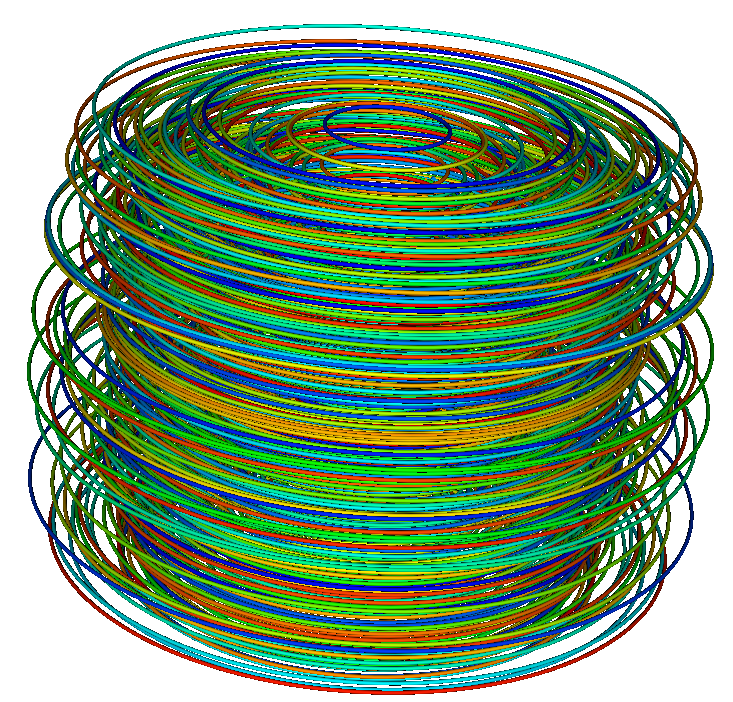}}
\caption{Images of streamlines generated from the five data sets used in this study.}
\vspace{-5pt}
\label{fig:datasets}
\end{figure*}


Our workloads are varied in the flow dataset, the amount of partitioning and parallelism, and the maximum iteration length. The following five datasets with different flow characteristics are used. 
\begin{itemize}
  \item \emph{Tokamak}: a torus-shaped vector field from a magnetically confined fusion simulation using the NIMROD code~\cite{sovinec2004nonlinear}. See Figure~\ref{fig:fusion}.
  \item \emph{Supernova}: a vector field derived from the magnetic field during a supernova simulation using the GenASiS code~\cite{Endeve_2010}. See Figure~\ref{fig:astro}.
  \item \emph{Hydraulics}: a thermal hydraulics simulation of a box with two inlets and one outlet using the NEK5000 code~\cite{fischer2008petascale}. See Figure~\ref{fig:fishtank}.
  \item \emph{CloverLeaf3D}: a hydrodynamics simulation of directional energy confined in a box using the CloverLeaf3D code~\cite{cloverleaf}.  See Figure~\ref{fig:cloverleaf}.
  \item \emph{Synthetic}: a perfectly cylindrical flow around a central axis. See Figure~\ref{fig:syn}.
\end{itemize}

Each dataset is sampled onto a sequence of five multi-block datasets containing 8, 16, 32, 64 and 128 blocks, respectively.
Each block within the multi-block dataset is a uniform grid of dimension $128 \times 128 \times 128$. The goal of data re-sampling operation is to prepare datasets used for different experiment configurations (8 to 128 MPI ranks with one data block per rank).
The workload for each configuration consists of 5000 particles placed randomly inside each block.
An overview of the block configuration is given in Table \ref{tab:block-layout}.

\begin{table}[tb]
\caption{Block layout configurations.}
\label{tab:block-layout}
\centering
\begin{tabular}{rccr}
\toprule
\multicolumn{1}{c}{\textbf{Num}} & \textbf{Block} & \textbf{Total} & \multicolumn{1}{c}{\textbf{Total}} \\
\multicolumn{1}{c}{\textbf{Ranks}} & \textbf{Layout} & \textbf{Mesh Size} & \multicolumn{1}{c}{\textbf{Particles}} \\
\midrule
8                  & $2 \times 2 \times 2$                 & $256 \times 256 \times 256$        & 40,000                 \\
16                 & $2 \times 2 \times 4$                 & $256 \times 256 \times 512$        & 80,000                 \\
32                 & $2 \times 4 \times 4$                 & $256 \times 512 \times 512$        & 160,000                \\
64                 & $4 \times 4 \times 4$                 & $512 \times 512 \times 512$        & 320,000                \\
128                & $4 \times 4 \times 8$                 & $512 \times 512 \times 1024$       & 640,000                \\



\bottomrule
\vspace{-10pt}
\end{tabular}
\end{table}

\begin{figure*}[ht]
\centering
\includegraphics[width=181mm]{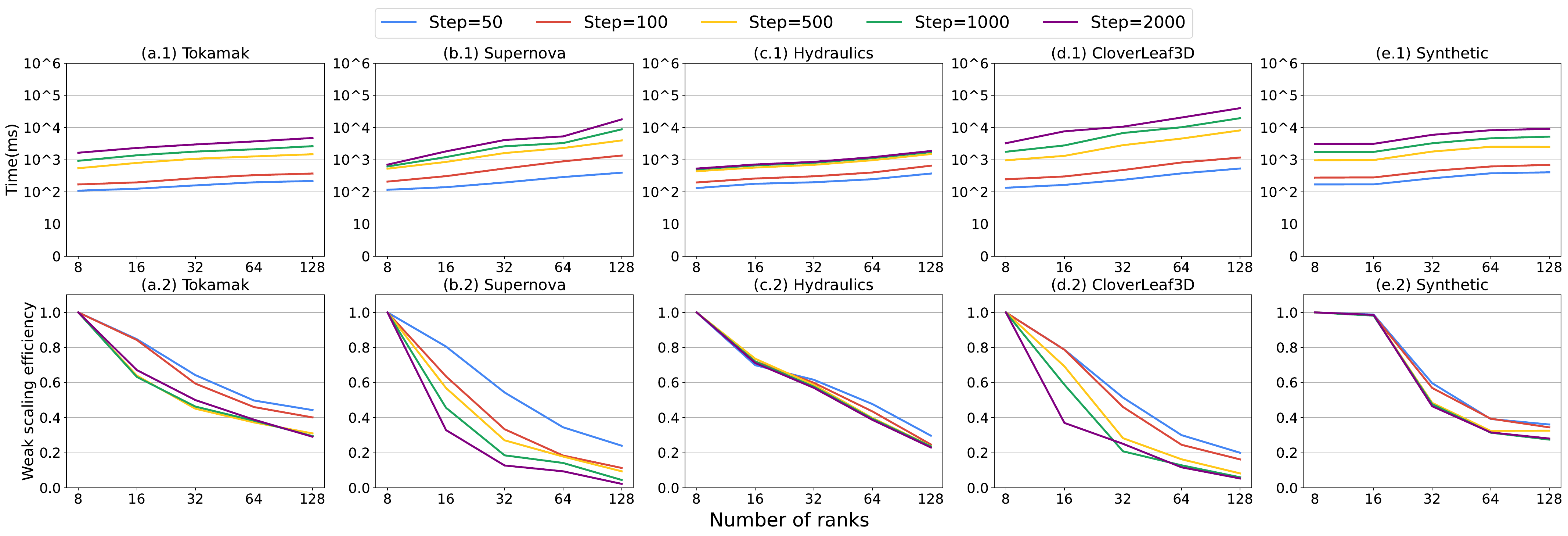}
\caption{POD particle advection as parallelism increases, organized by dataset.
The top row shows the execution time of the particle advection algorithm,
while the bottom row shows weak scalability.
The X-Axis for all sub-figures is the number of ranks ($log_2$ scale).
The Y-Axis for the top row is execution time ($log_{10}$ scale) and for the bottom row is efficiency relative
to the $8$-rank case.
}
\label{fig:results:scale}
\vspace{-10pt}
\end{figure*}

Particles are traced using five different maximum iterations: $50$, $100$, $500$, $1000$, and $2000$ steps.
Overall we test ($5$ datasets $\times 5$ levels of parallelism $\times 5$ different advection steps) = $125$ total configurations. These configurations cover the different use cases for flow visualization; for example, FTLE may use short iterations such as $50$ whereas streamline is more likely to use longer iterations such as $2000$. For the implementation, we use the particle advection in VTK-m~\cite{Pugmire2018}, which supports different types of single-core and multi-core backends.
The focus of this paper is to identify the underlying causes of workload imbalance in the POD particle advection algorithm. To simplify our analysis, we use the serial backend (i.e., each MPI process uses a single CPU core) in our experiments to eliminate effects caused by shared-memory architecture. The use of GPUs may introduce additional costs that compound the underlying performance issues~\cite{Childs2014}. Therefore, we believe configurations shown in Table \ref{tab:block-layout} represent reasonable possible cases for measuring the MPI-parallel efficiency of POD particle advection algorithm in this paper. 

The runs are performed on the Summit supercomputer at Oak Ridge National Laboratory~\cite{summit}.
Summit is a 200 PetaFLOPS supercomputer that consists of 4608 nodes. Each node contains two 22-core IBM Power9 CPUs and six NVIDIA Tesla V100 GPUs, and a Mellanox EDR 100G InfiniBand network~\cite{Vazhkudai2018}.

\subsection{Analysis data collection}
\label{subsec:exp_data_collection}
The code described in Algorithm~\ref{pvalgo} is instrumented with a number of timers and counters, which are used to collect data from both a rank-centric and particle-centric perspective.
For the rank-centric analysis, timers and counters are used to collect data associated with each sub-task within the algorithm. These sub-tasks are \ModelIWord{\emph{Initialization}}, \ModelBOWord{\emph{Begin Overhead}}, \ModelAWord{\emph{Advection}}, \ModelEOWord{\emph{End Overhead}} and \ModelCWword{\emph{Communication}}.
The timers record the time spent in each phase and the counters quantify such things as the total number of rounds, number of particles processed, number of particles sent and received, etc.

For particle-centric analysis, timing and counter attributes are added to each particle.
These timers and counters are updated during the execution of the algorithm.
The timers record the accumulated time spent in each of the sub-tasks in the algorithm as well as the time of termination.
The counters record the number of traversed blocks, number of communications, and number of advection steps. The reason for particle termination is also recorded.
These values are held in memory during algorithm execution, aggregated, and output after completion.

We also provide the ability to specify a unique particle ID to collect detailed information for a single particle.
In this mode, each rank checks the particles being processed and records additional information if the specified particle ID is found. In this way, we can trace detailed information about a specific particle, such as which ranks are traversed and associated timestamps when a particle is sent and received. Using this capability we are able to generate detailed information on the long running particle in a given workload.
The long running particle is determined by running a test case and recording the termination time for each particle. The test case is then re-run using the ID for the long running particle to collect particle-specific information.\footnote{In the second run, the particle with the specified ID may not technically be the longest one because of uncertainty in communication, but it provides a representative long-running particle accurately.}
The experiments are run both with and without collecting this data to verify that our instrumentation does not significantly alter runtime behavior. We find that the overhead for collecting this data is at most $2.8\%$.

\section{experimental results}
\label{sec:expresults}






Experimental results are divided into four sections:
\S\ref{sec:expresults:exectime} considers overall execution time and weak scalability,
\S\ref{sec:expresults:rank} and \S\ref{sec:expresults:slowparticle} 
analyze performance based on the rank participation and the particle-centric model,
respectively, and
\S\ref{sec:expresults:allparticles} analyzes termination behavior.

\subsection{Execution time analysis}
\label{sec:expresults:exectime}


Our experiments' execution times are summarized in Figure~\ref{fig:results:scale}.
The first row in Figure~\ref{fig:results:scale} shows the total execution time for a weak scaling run on each of the five datasets, with $25$ configurations 
for each dataset ($5$ levels of parallelism $\times$ $5$ values for maximum number of advection steps).
As described in the experimental setup (\S \ref{subsec:exp_setup}), we maintain a fixed count of $5000$ seeds per data block. 
We also assign one block per rank, so the total number of ranks and blocks is the same. 
The weak scaling efficiency for each experiment is shown in the second row of Figure~\ref{fig:results:scale}.
We use the runtime with $8$ ranks (the lowest level of concurrency in our experiments) as the baseline, i.e.,
if $T(N)$ is the runtime with $N$ ranks, then the weak scaling efficiency for $N$ is $T(N)/T(8)$.
This is consistent with a unit-cost-based efficiency~\cite{Moreland2015:ISC} using 8 ranks as a base case and recognizing that the problem size increases proportionally to the number of ranks.
For this formulation, perfect scaling would have an efficiency value of $1$ for all numbers of ranks.
That said, for our experiments, the weak scaling efficiency significantly declines as the number of ranks increases, even for cases that are well suited for the POD algorithm (i.e., Tokamak and Synthetic).
Additionally, the bottom half of Figure~\ref{fig:results:scale}
shows that increasing the number of advection steps results in decreased efficiency, i.e.,
the yellow, green, and purple lines ($500$, $1000$, and $2000$ advection steps) have the worst
efficiency, in particular with the Supernova and the CloverLeaf3D datasets.


\subsection{Rank participation}
\label{sec:expresults:rank}
\begin{figure*}[ht]
\centering
 
\includegraphics[width=173mm]
{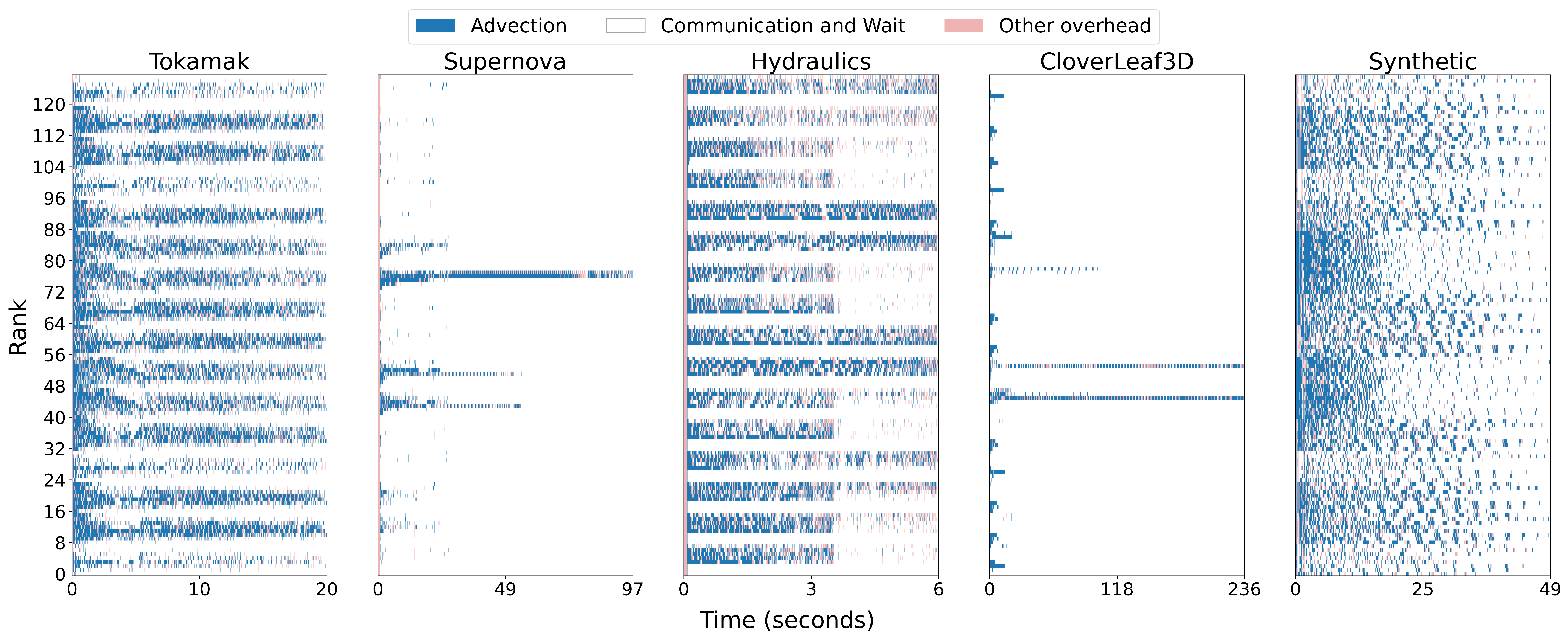}
\caption{Gantt charts for experiments consisting of 128 ranks and 2000 advection steps. 
Each chart shows the activity for each rank over the course of the run:
blue regions denote advection time, white regions represent both communication time and wait time, and pink regions represent other overheads.}
\label{fig:results:gantt}
\vspace{-5pt}
\end{figure*}

Gantt charts provide an effective way to understand the activity of each rank over the course of the execution.
Figure~\ref{fig:results:gantt} shows Gantt charts for a workload consisting of 128 ranks and 2000 steps for each dataset.
These charts provide an overview of the balance of work among the ranks and give insight into the overall scalability. 
The Gantt charts of the Tokamak, the Hydraulics, and the Synthetic datasets display a more even distribution of workloads. In contrast, the Gantt charts of the Supernova and the CloverLeaf3D datasets reveal that a few ranks are tasked with much more work, leading to poor efficiency and longer execution times.

To evaluate the overall efficiency of the algorithm during it's execution, we use the rank participation metric introduced in \S~\ref{subsec:model:metrics}.
%
A rank participation of $1$ indicates that the workload is perfectly balanced among all ranks whereas a rank participation of $1/N$ suggests that only one rank is working while the others remain idle.



\begin{figure*}[ht]
\centering
\label{fig:scale}  \includegraphics[width=180mm]{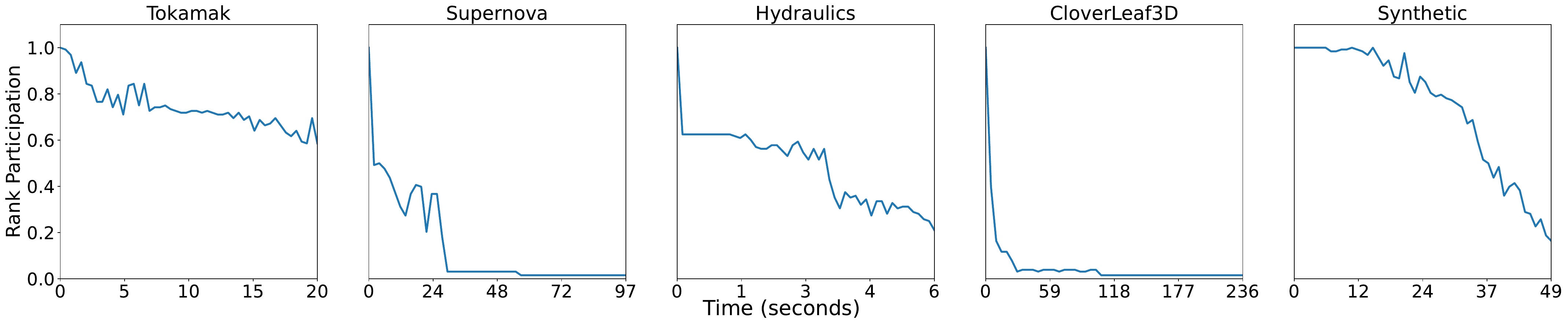}
\caption{Rank participation values of all evaluated datasets based on experiment results from Figure~\ref{fig:results:gantt}. The $x$ axis represents the execution time of the workload, and the $y$ axis represents the corresponding rank participation value at each moment. 
}
\vspace{-5pt}
\label{fig:results:pr}
\end{figure*}


Figure~\ref{fig:results:pr} shows the rank participation as a function of time for the same runs from Figure~\ref{fig:results:gantt}, i.e., those with $128$ blocks and $2000$ advection steps.
The rank participation curve provides a straightforward view of the number of ranks actively engaged as the algorithm progresses. We note that a rank is marked as participating when it is performing any of the following tasks from Equation~\ref{model:eq1}:
{\ModelA}, {\ModelBO}, or {\ModelEO}.
And conversely, a rank is marked as not participating when performing tasks \ModelC\ or \ModelW\ from Equation~\ref{model:eq1}.

The rank participation charts shown in Figure~\ref{fig:results:pr} highlight the problems with efficiency for the Supernova and the CloverLeaf3D datasets, as there is a dramatic drop in participation at the outset of the execution.
The Tokamak and the Synthetic datasets exhibit relatively consistent and steady participation rates over the course the execution while the Hydraulics dataset lies somewhere in the middle.

\begin{figure*}[ht]
\centering

\includegraphics[width=180mm]{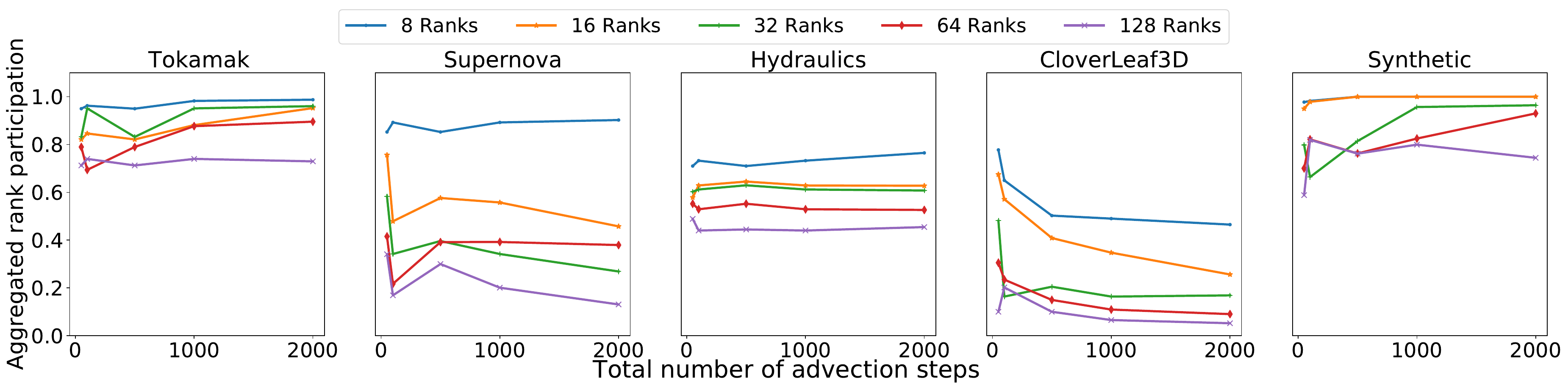}
\caption{Aggregated participation for all experiments.  The five subplots correspond to the five datasets, and each colored line within a subplot corresponds to the number
of ranks.  Note that the X-Axis is showing behavior as the number of advection steps to take increases, i.e., the tick marks at 1000 correspond to the behavior
across an entire experiment that has particles travel 1000 steps, while the tick marks at 2000 correspond to different experiments where the particles
travel 2000 steps.
}

\label{fig:summary_RP}
\vspace{-2pt}
\end{figure*}

We further compute the \emph{aggregated rank participation} over the course of a run by integrating the participation rank curve (the area under the rank participation curve) across a normalized time axis. If all ranks were engaged across the entirety of the run, the aggregated participation rate would be $1$. 
The aggregated rank participation values are shown in Figure~\ref{fig:summary_RP} for all of the configurations evaluated in our study. 
The curves in each subplot track the aggregated rank participation for a run at a given level of parallelism as a function of the number of advection steps.
As an example, the purple line in the Tokamak subplot shows the aggregated rank participation running on $128$ ranks and advection steps of 50, 100, 500, 1000, and 2000.

Plotting these metrics provides insight into the scalability of the algorithm across different datasets and workloads. 
Specifically, the Tokamak and the Synthetic data have high average rank participation values, whereas the Supernova and the CloverLeaf3D exhibit lower participation values when the number of ranks increases from $8$ to $128$. The average participation rates for the Supernova and the CloverLeaf3D datasets are more sensitive to the number of ranks compared to other datasets. 

\subsection{Particle-centric model}
\label{sec:expresults:slowparticle}

\begin{table*}[ht]
\begin{center}
\vspace{-3pt}
\caption{
Key statistics of long running particles for evaluated datasets. $T_p$ represents the total time of long running particle. \BOcolor{$BO$} and \EOcolor{$EO$} represent the percentage of begin overhead and end overhead, respectively. \Acolor{Advection} represents the percentage of advection time. \CWcolor{``Comm''} represents the percentage of time spent sending particles between ranks, and \CWcolor{``Wait''} represents the percentage of wait time spent in \CWcolor{"Comm and Wait"} toward the total time. $N_p$ represents the number of times the long running particles traveling through ranks.
}
\label{tab:long_particle}
\begin{tabular}{lccccccr}
\toprule
Dataset & $N_p$ & \BOcolor{$BO$} & \EOcolor{$EO$}  & \Acolor{Advection} &  \CWcolor{Comm} & \CWcolor{Wait} & $T_p$ (seconds) \\
\midrule
Tokmak      & 122                 & 4\%                 & 14\%                & 54\%          & 2\%  & 26\%               & 19.4                         \\
Supernova    & 275                 & 6\%                 & 21\%                & 58\%         & 3\%  & 12\%              & 97.2                         \\
Hydraulics    & 213                 & 5\%                 & 11\%                & 45\%           & 2\%  & 37\%               & 5.9                          \\
CloverLeaf3D             & 186                 & 6\%                 & 21\%                & 59\%         & 3\%  & 12\%              & 235.6                        \\
Synthetic                & 192                 & 5\%                 & 21\%                & 58\%       & 4\%  & 12\%              & 46.8                         \\ 
\bottomrule
\end{tabular}
\end{center}
\vspace{-10pt}
\end{table*}


As described in the model for execution time (\S\ref{subsec:model:def}), the overall execution time for the particle advection algorithm is only as fast as the slowest
particle, i.e., the particle that takes the longest
to complete execution.
This section focuses on the slowest particle for the configurations with 128 ranks and $2000$ steps for each of our five datasets.
%
The results are shown in Table~\ref{tab:long_particle}. 
%
Of note, the slowest particle is non-deterministic, as
the time it takes for a given particle to complete can change slightly from
run to run, meaning the slowest in one experiment may get edged out by
another particle in the next experiment.
For these results, we ran an experiment to identify the slowest particle for that
run,
and then re-ran the experiment with diagnostic information for that particle.
We continue to refer to this particle as the ``slowest particle'' in the second experiment.

Table~\ref{tab:long_particle} is organized according to Equation~\ref{model:eq2}. 
This table reveals three interesting observations.
%
%
%
%
First, the variation in $T_p$ varies considerably.
The reasons behind this observation are explored in \S~\ref{sec:variation}.
Second, ``end overhead'' (\EOcolor{$EO$}) is taking a significant portion of the overall execution time, which is explored in \S~\ref{subsec:eo}.
Finally, the number of MPI ranks visited ($N_p$) is quite large,
indicating that these particles are bouncing between
MPI ranks.
\S~\ref{subsec:pingpong} analyzes the reason,  and describes a phenomenon we refer to as ``\textbf{ping pong particles}.'' 

\begin{figure*}[ht]
\vspace{-5pt}
\centering
\includegraphics[width=170mm]{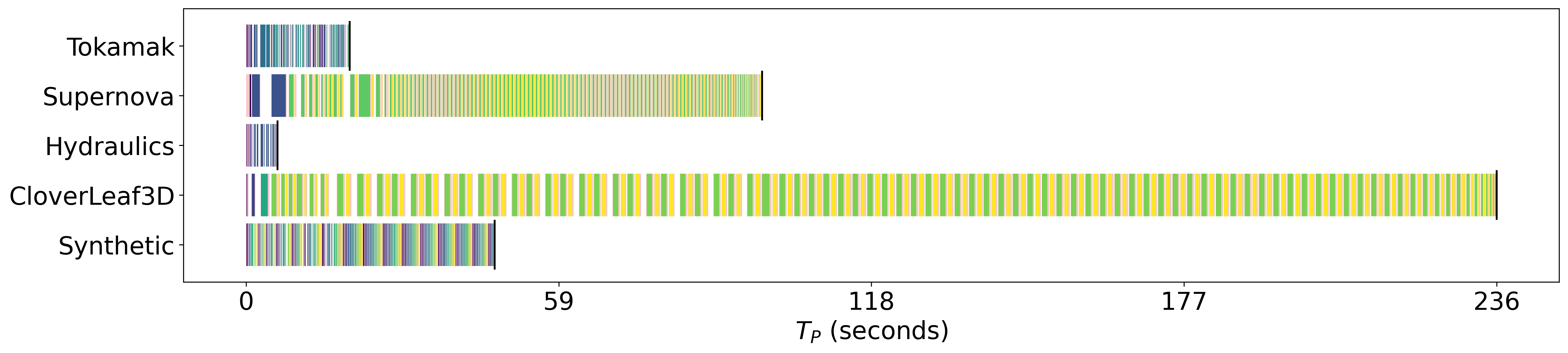}
\caption{Gantt charts of the particle with the longest execution time for experiments where there are $128$ blocks and $2000$ advection steps. 
The white color bars represent the communication and wait stage of the longest running particle. The pink color represents {\ModelBO}, {\ModelEO}, {\ModelI}, and other uncategorized time between measurements of the longest running particle. Other colored bars represent the \Acolor{advection} time, with each color representing time spent
on a different MPI rank.  For example, for CloverLeaf3D,
the particle travels back and forth between two MPI ranks
colored yellow and green. 
Finally, the end of each experiment (i.e., the termination
of the longest running particle) is indicated with
a vertical black line.
}
\label{fig:particle_long_exec}
\vspace{-6pt}
\end{figure*}

\begin{figure*}[ht]
\centering
\includegraphics[width=180mm] {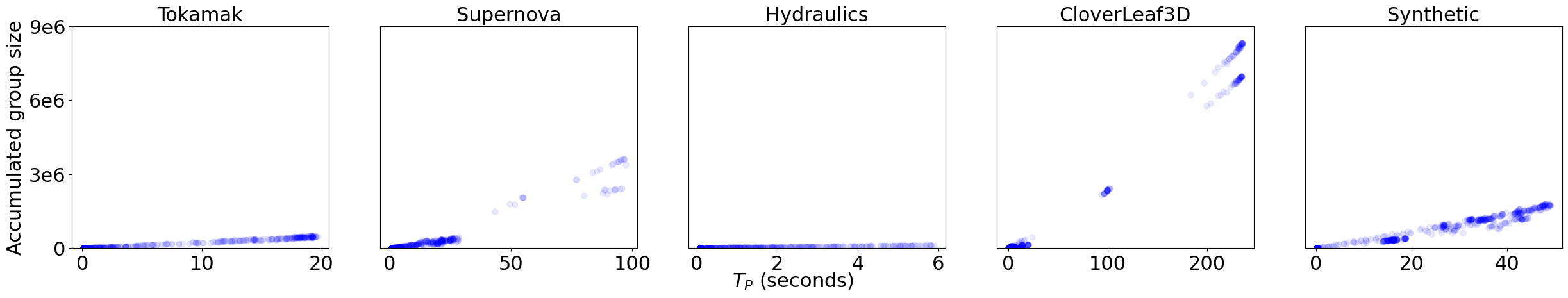}
\caption{Scatter plots for particles for the experiments with 128 blocks and 2000 advection steps. The $X$ axis is the execution time for each particle ($T_p$) and the $Y$ axis is the accumulated group size for the particle as it traverses through blocks. The particles plotted are sampled at a rate of one of out of every $1000$. 
}
\label{fig:scatter_accGroup_aliveTime}
\vspace{-10pt}
\end{figure*}

\subsubsection{Variation in $T_p$}
\label{sec:variation}
$T_p$, the time it takes for a particle to complete, is the sum of all of the factors in Equation~\ref{model:eq2}. 
Despite the longest-running particles having the same amount of work to do as the
other particles, their lifetime varies
tremendously, from 5.9s for the Hydraulics dataset to 235.6s for
the CloverLeaf3D dataset.
%
%

The variation in execution time is due to the load on a given MPI rank.
When a particle arrives at an MPI rank, it is processed as part of a group.
If the group is large, then the advection time for that particle on that rank will take
longer.
For CloverLeaf3D, the group sizes get as large as 100,000 particles, while for Hydraulics they are
often under 10,000.
Figure~\ref{fig:particle_long_exec} shows a Gantt chart for the slowest particles.
For CloverLeaf3D, the green and yellow rectangles 
are as wide as they are
because advection time covers not only the slowest particle,
but also all of the particles it arrived with.
%
%
%
So while a given MPI rank is working diligently on advecting particles,
these particles are taking long periods to move on to the next rank since there are so many particles to process.
Further, the issue is not that the particles are being processed in atomic
groups;
we experimented with communication schemes that sent smaller batches to other
MPI ranks when they were completed and found the overall execution time did 
not change significantly.
%
%
Figure~\ref{fig:scatter_accGroup_aliveTime} shows the delays resulting from large group sizes
in another way.
For a given particle, it shows the ``accumulated group size,'' i.e., total group size over all the MPI ranks it visited.
For example, if a particle visited three MPI ranks and was part of groups of size 100, 200, and 600,
then its accumulated group size would be 900.
This plot shows that CloverLeaf3D's slowest particle has an accumulated group size of almost nine million, while
Hydraulics is well only around $0.1$ million, correlating closely with $T_p$.

\subsubsection{End overhead}
\label{subsec:eo}

End Overhead (\ModelEO) is surprisingly larger than Begin Overhead (\ModelBO), taking
$2\times$--$4\times$ more time.
Furthermore, the \ModelEO\ time tends to grow with the size of the particle group being processed.
This is because there is significant bookkeeping at the end of the advection round to manage the resulting particle states, which can be different for each particle.
The particles must be processed to determine which are active and which are terminated.
Those that are active must be partitioned based on the block to which they will enter, and their respective data must be placed in buffers to be ready for the communication step.
We highlight this result as we note previous work has (reasonably) focused on optimizing the computation directly involved in \Acolor{advection}, but the time for auxiliary computation is still important.
For a single-block the effect of \EOcolor{$EO$} is insignificant because there is less bookkeeping.
But with MPI communication, this process has to happen regularly as particles pass in and out of blocks.
%
%

\begin{figure}[ht]
\centering
\subfigure{
\includegraphics[height=30mm]{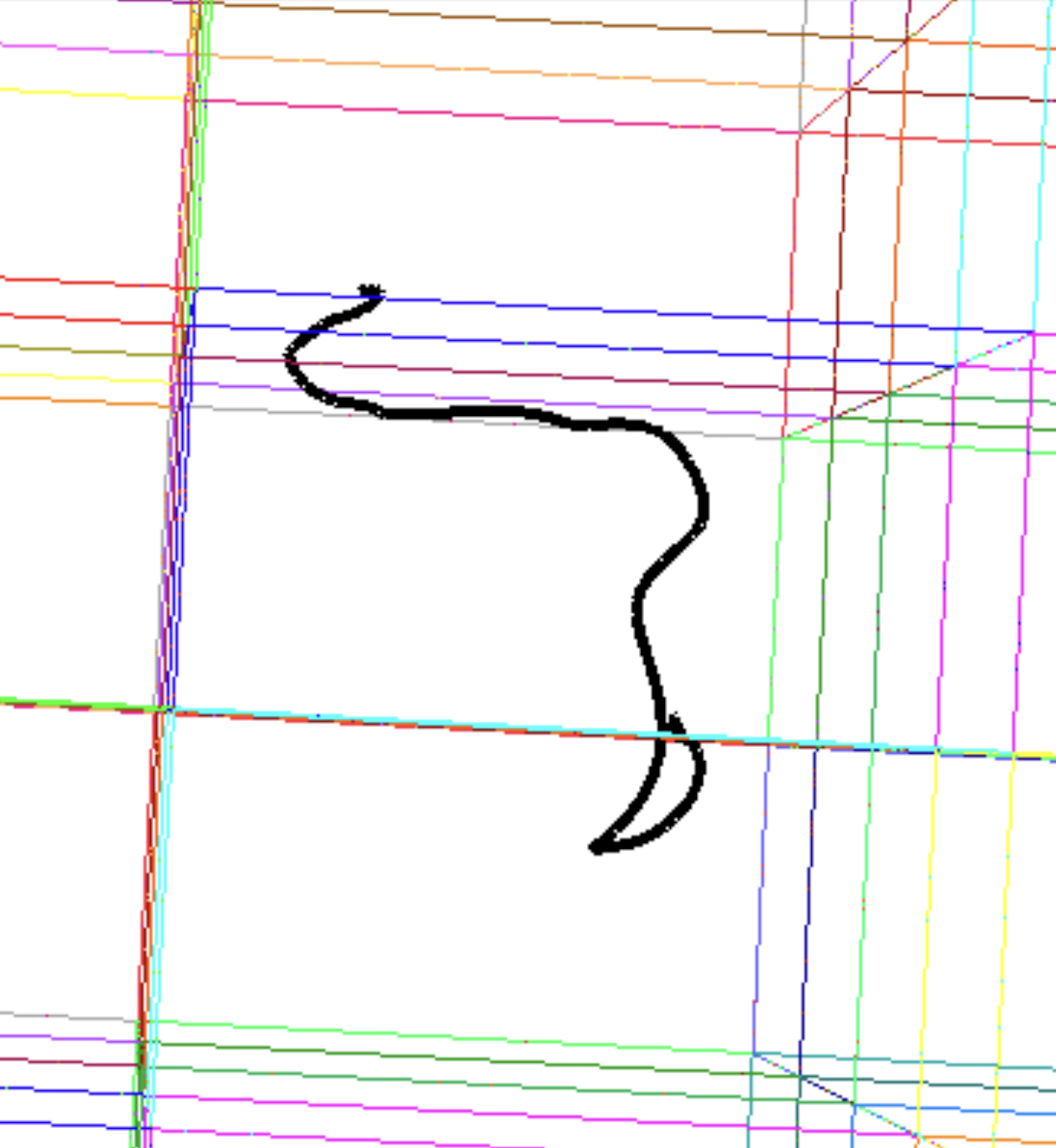}
}
\hspace{5mm}
\subfigure{
\includegraphics[height=30mm]{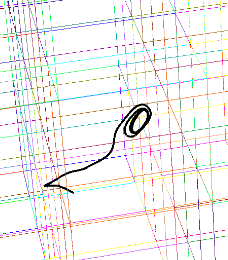}
}
\caption{Examples of particles that ping pong in the Supernova (left) and Cloverleaf3D (right) datasets. The long running particles are trapped into a vortex crossing two adjacent data blocks.}
\label{fig:example_pingpong}
\vspace{-10pt}
\end{figure}

\subsubsection{Illustration of ping pong particles}
\label{subsec:pingpong}

Figure~\ref{fig:particle_long_exec} shows the pattern followed by the slowest particle
in each configuration.
In all cases, the amount of solid color (processing
by an MPI rank) is greater than the amount of white (idle time between MPI ranks),
which is consistent with Table~\ref{tab:long_particle}.
For the CloverLeaf3D dataset, the long particle spends most of its time back
and forth between the ``green'' and ``yellow'' MPI ranks.
We refer to particles that repeatedly pass back and forth between two adjacent MPI ranks as ``ping pong particles.''
These particles suffer from both of the issues just identified ---
they spend more time waiting in queues to advect and they spend more time on end overhead.
The fundamental issue is not that they are traveling back and forth between the same two blocks, but rather that they complete so few
advection steps before moving on to the next block, and thus need to visit many blocks and incur extra overheads.

For the Tokamak and the Synthetic datasets, although the particle travels in a loop, there are more than two MPI ranks involved along the particle advection path.
For the Supernova and the CloverLeaf3D datasets, the stage where the particle circulates between ranks takes more than half of the total execution time.
Particles exhibiting this behavior are shown in Figure~\ref {fig:example_pingpong}. In particular, a vortex that is near block boundaries exists in both datasets, resulting in particle circulation between two blocks.
We use the term ``ping pong effect'' to describe that a large number of ping pong particles are circulating in a vortex that spans MPI
ranks and thus take a comparatively long time to complete their advection step. 
\S~\ref{discus:pingpong} provides a more detailed analysis of ping pong effects, as well as considering approaches to improve execution
time.
%

\subsection{Overview of statistics for terminated particles}
\label{sec:expresults:allparticles}
This subsection presents statistics collected for all particles, which
contrasts with the previous subsection's focus on long-running particles.
During each run, the following data are collected for each particle: the time when a particle terminates, the number of blocks traversed by the particle, and the reason the particle terminated. In particular, reasons for termination include:
\begin{itemize}
    \item Out of bounds: The particle exited the spatial bounds of the dataset.
    \item Zero velocity: The particle enters a sink.
    \item Max step: The maximum number of advection steps is achieved.
\end{itemize}

\begin{table}[ht]
\begin{center}
\vspace{-5pt}
\caption {The percentage of particles terminated for each of the three termination criteria. This data come from the experiments with $128$ data blocks and $2000$ advection steps.}
\label{tab:particle_exit}
\begin{tabular}{llll}
\toprule
Datasets &  Out of bounds & Zero velocity &  Max steps  \\
\midrule
Tokamak  & $2.1\%$  & $69.1\%$ & $28.8\%$  \\
Supernova  & $82.5\%$  & $8.5\%$ & $9.0\%$  \\
Hydraulics  & $39.6\%$  & $59.0\%$ & $1.4\%$  \\
CloverLeaf3D  & $15.1\%$  & $41.9\%$ & $43.0\%$  \\
Synthetic  & $26.2\%$  & $12.7\%$ & $61.1\%$  \\
\bottomrule
\end{tabular}
\end{center}
\end{table}

\begin{figure*}[ht]
\centering
\includegraphics[width=178mm]{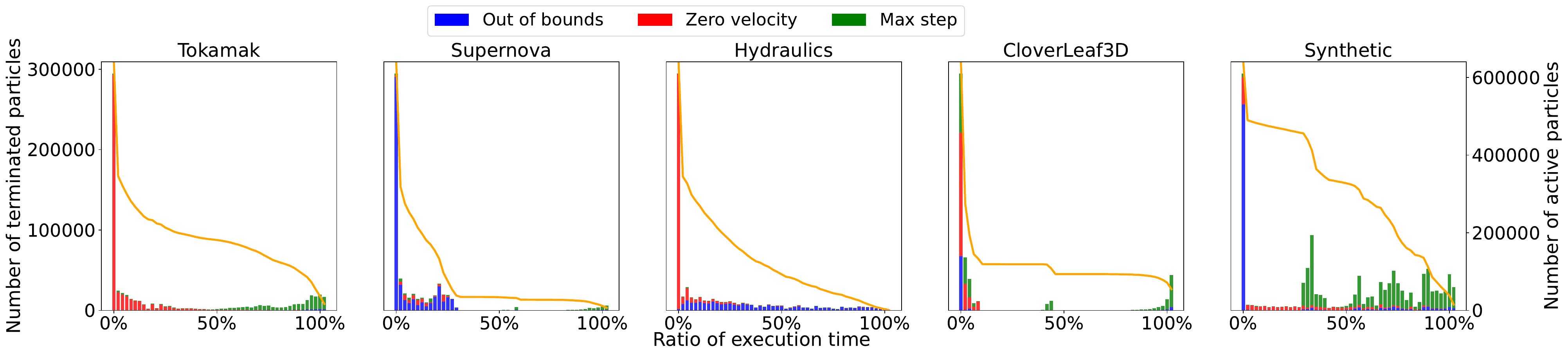}
\caption{The colored bars represent the number of terminated particles (labeled on the left $y$ axis) when using $128$ ranks and $2000$ advection steps. Different colors represent different termination reasons. 
The orange curve represents the number of active particles (labeled on the right $y$ axis) within a specific time slot during the particle execution.}
\label{fig:particle_num}
\vspace{-10pt}
\end{figure*}

Table~\ref{tab:particle_exit} lists the breakdown of reasons for termination across the evaluated datasets. Figure~\ref{fig:particle_num} further illustrates the number of terminated particles over time. Specifically, the colored bars represent the number of terminated particles for each time slot, and the orange colored curve represents the number of active particles. The results illustrated in Table~\ref{tab:particle_exit} and Figure~\ref{fig:particle_num} supplement the findings illustrated in Figure~\ref{fig:particle_long_exec}, further illustrating the variation in particle advection execution time for different datasets. 

One important observation is that the orange curve in Figure~\ref{fig:particle_num} shows a similar pattern for the Supernova and the CloverLeaf3D datasets. Both of them drop dramatically at the beginning of the algorithm execution and keep steady subsequently. However, only around $9\%$ of particles for the Supernova dataset complete all advection steps and terminate because of ``Max steps'' whereas there are $43\%$ particles for the CloverLeaf3D dataset (as shown in Table~\ref{tab:particle_exit}). Although most of these particles are ping pong particles, the large number of particles in the CloverLeaf3D dataset causes more overhead in the stage of traversing particles between different blocks compared to the Supernova dataset. As illustrated in Figure~\ref{fig:particle_long_exec}, the yellow colored region is wider in the CloverLeaf3D dataset than in the Supernova dataset.

Although the number of active particles shown in Figure~\ref{fig:particle_num} decreases gradually for the Tokamak, the Hydraulics, and the Synthetic datasets,\footnote{There are no ping pong particles for these datasets.} the termination reasons are distributed differently. For the Synthetic dataset, most green colored bars are located at the right half of execution time, indicating more particles are advected to maximal steps. In particular, Table~\ref{tab:particle_exit} illustrates $61.1\%$ of particles terminate because of ``Max steps'' for the Synthetic dataset. In contrast, there are only $28.8\%$ and $1.4\%$ for Tokamak and Hydraulics, respectively. The percentage of particles terminated by completing maximal advection steps also corresponds to the particle advection execution time shown in Figure~\ref{fig:particle_long_exec} where the Hydraulics dataset has the shortest particle advection time and Tokamak and Hydraulics datasets take a little bit longer.

\section{Discussion}
\label{sec:discussion}

In this section, we discuss two significant issues with parallel particle advection. 
The first is the surprisingly low participation rate of ranks in a parallel job, even for use cases where the POD algorithm is well suited.
The second is our observation of the ``ping pong effect'' where the runtime can be extended by ``ping pong particles.'' These particles are exchanged back and forth among a small number of processes (such as two adjacent blocks explored in this paper) as their trajectory loops between them.
The ping pong effect happens more often and with more particles than we were expecting. Although the rank participation and ping pong effect are inextricably linked, we discuss them separately to address the features specific to each in \S~\ref{discus:rankp} and \S~\ref{discus:pingpong}, respectively. 
Finally, we discuss several potential solutions to resolve the issue caused by ping pong particles in \S~\ref{discus:potential_solutions}.


\subsection{Rank participation}
\label{discus:rankp}
The potential for load imbalance for POD particle advection has long been known.
However, the actual effect of this load imbalance has been anecdotal at best and is often discounted as too minor or too rare of an effect to justify an implementation in production tools.
In our quantitative study of load balance through rank participation, we find that load balance is worse than expected under conditions surmised to be favorable (e.g., the Tokamak and Synthetic datasets where the flow continuously circulates).

As mentioned in the related work (\S~\ref{sec:related-work}), previous investigations provide techniques, particularly hybrid POD/POS techniques \cite{Pugmire2009,kendall2011simplified,Peterka2011:IPDPS,guo2013coupled, binyahib2019lifeline,binyahib2021hylipod}, that should improve the load balancing, and we should investigate integrating these optimization mechanisms into production tools.
That said, the problem becomes even more challenging for on-node \emph{in situ} visualization \cite{InSituTerminology} where mesh data placement is predetermined and mesh data movement is restricted or infeasible.

The load balancing issue suggests that off-node \emph{in situ} (often referred to as \emph{in transit}) would provide a double win over on-node \emph{in situ}.
First, placing the visualization on a job with fewer processes can improve tasks with scaling or load balancing issues \cite{Kress2019,kress2020opportunities}.
Second, because the data are moved from one set of nodes to another, a repartition of the data can happen to help re-balance the work. However, these strategies require workload anticipation based on prepossessing~\cite{nouanesengsy2011load} or information collected in real time~\cite{zhang2018dynamic,xu2022reinforcement}.
\subsection{Particle group size and the ping pong effect}
\label{discus:pingpong}

The authors have long suspected the possibility of particles rapidly passing back and forth between partitions, but never previously identified a case.
The use and analysis of Gantt charts discussed in \S\ref{sec:expresults} proved invaluable in identifying and understanding these cases and the rank participation analysis helped to quantify their impact.

We are surprised to note that the ping pong effect is not caused by a lone particle but rather by a large group of particles.
Consider the Gantt chart for the CloverLeaf3D data in Figure~\ref{fig:results:gantt}, which shows the majority of execution time spent in ping pong particles between two ranks (specifically 45 and 53).
We can further see from Figure~\ref{fig:particle_num} that this is not a single particle but around 100,000 particles. 
In fact, the ping pong effect only becomes an issue when large numbers of particles are involved.

From the perspective of a particle-centric model (Equation~\ref{model:eq2}), 
a particle $p$ passing through block $i$ has a minimal value of \pModelA\ if it is being processed alone.
However, in practice, a particle is bundled with other particles that have arrived at a given block.
Such behavior is desirable to keep processors busy, but
\pModelA\ grows as particle $p$ must wait for its fellow particles to also complete before it can move to the next processor.
This delay can compound the value of $\sum_i$ \pModelA.


Additionally, when a group of particles traverses a large number of blocks, these particles will incur the extra overheads 
(\pModelBO\ and \pModelEO) associated with entering and exiting a block.
Paths traversing many blocks also incur more overall communication and wait time (\pModelCW).
%
%
%
A large variation of $T_p$ among particles indicates an imbalance of particle advection in the system.
When a particle has a large batch size among its advection path and a long traversing block sequence at the same time, its execution time will be significantly larger than other particles.
Avoiding these long-running particles can contribute to a more balanced particle advection system and decrease the total execution time.

Looking again at the CloverLeaf3D Gantt chart in Figure~\ref{fig:results:gantt} we see a second pair of ranks with a ping pong effect (between rank $77$ and rank $78$). 
These particles finish faster because they are in smaller groups.
This smaller group size can be observed by the smaller dip of active particles about midway through the CloverLeaf3D chart of Figure~\ref{fig:particle_num}.
These results illustrate that both large group size and ping pong particles compound to worsen the execution time.

The fact that the ping pong effect worsens as more particles participate suggests that this is not caused by communication latency.
An obvious problem with a large number of ping pong particles is that a large number of long trajectories must be computed by a small number of processes.
However, the same could be said if a large number of particles have long, cyclic trajectories contained within a single partition, yet our measurements suggest that a ping pong effect is worse despite having two processes to divide the work.

The ping pong particles clearly have more overhead than the equivalent trajectory on a single partition.
One potential overhead is the time to transfer particle data between processes.
However, our profiling clearly shows transfer time to be insignificant.
Rather, we find overheads associated with entering and leaving mesh blocks to be substantial.
In particular, the \ModelEO\  time required to reorganize particles based on their termination status and destination is larger than expected and grows proportionally to the number of particles in the group.

This ping pong effect suggests a new area of investigation for particle advection on HPC systems.
Previous investigations have rightfully focused on optimizing the advection loop as most of the computation happens there.
However, we find that the ``insignificant'' overhead incurred during initialization and termination of the particles can accumulate greatly as particles pass across multiple block boundaries.
We therefore note the potential for improving large-scale particle advection by optimizing this part of the process.


\subsection{Potential solutions for avoiding long running particles}
\label{discus:potential_solutions}





The factors discussed in this paper about long-running particles can shed light on several potential solutions for improving the load balance and running time of particle advection.
We propose some hypotheses for potential future research that could improve the overall runtime of parallel particle advection by reducing the times of the longest-running particles.
To be clear, providing solutions to these problems is beyond the scope of this paper, but we provide some evidence to point to fruitful areas of research.

\begin{figure}[htb]
\centering

\includegraphics[width=73mm]{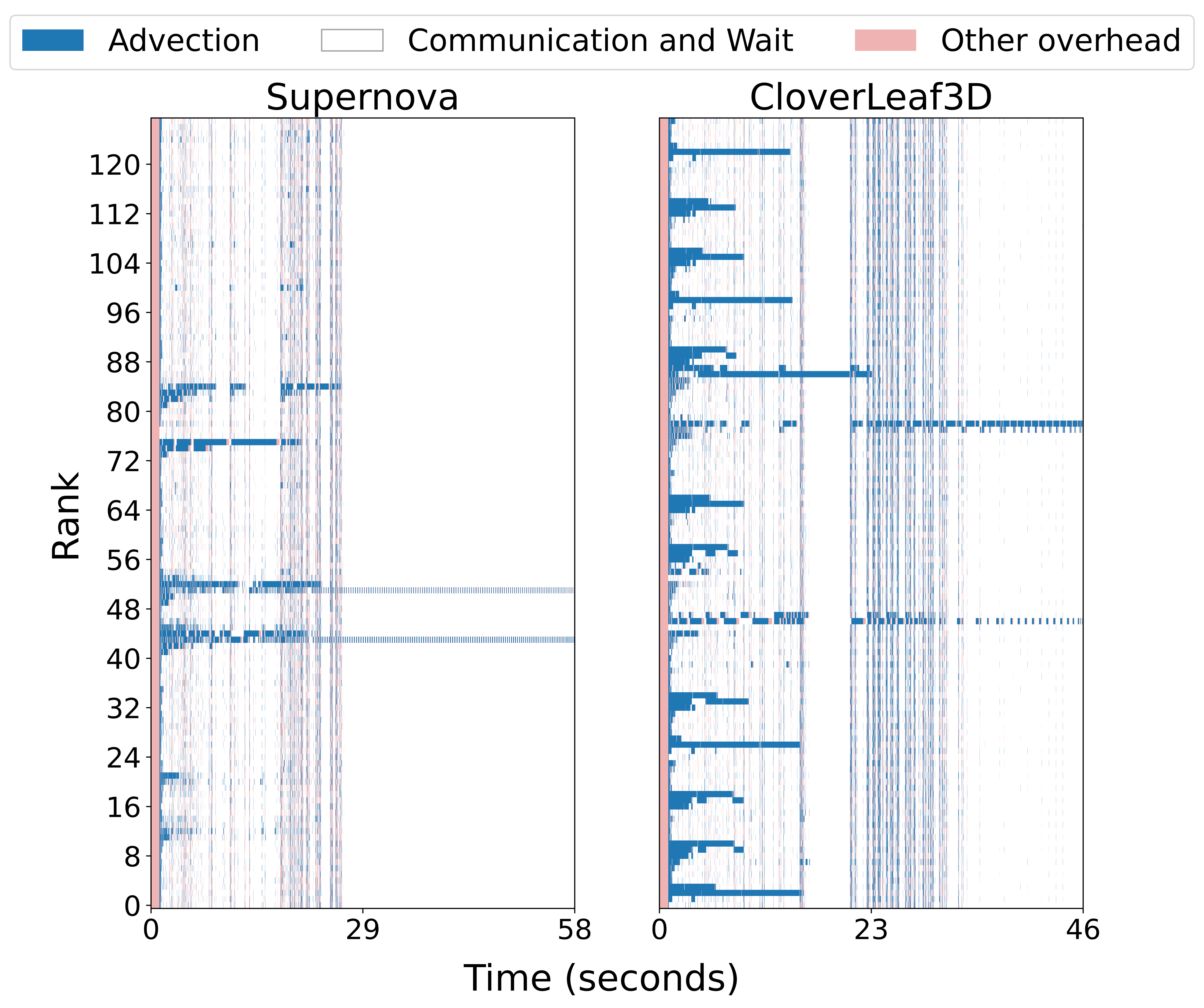}
\caption{Gantt charts of the experiments rerun with the blocks of ping pong particles \emph{duplicated} across all ranks.
    Supernova (left) has blocks 76 and 77 duplicated across all other ranks whereas Cloverleaf3D (right) has blocks 45 and 53 duplicated across all other ranks.} 
\label{fig:discuss:dup}
\vspace{-10pt}
\end{figure}

Consider that one of the problems with ping pong particles is that a small number of ranks do extra work with the ping pong overhead.
We hypothesize that we can increase load balance and reduce overall runtime by dividing this overhead among more processes.
This can be done by duplicating problematic blocks.
Figure~\ref{fig:discuss:dup} demonstrates this idea for the Supernova and the Cloverleaf3D datasets, the two datasets with the most pronounced ping pong effect.
Using our post hoc knowledge from the previous experiments, we duplicate the blocks from the two ranks having the longest running particles to all other ranks.
That is, each rank with ID $i$ will load the block with ID $i$ and one of these other two duplicated blocks.
Two observations can be made from Figure~\ref{fig:discuss:dup}.
First, we see that the ranks that previously had the longest running ping pong particles (76 and 77 for Supernova, 45 and 53 for Cloverleaf3D) are no longer ping pong particles and are no longer the bottleneck of the operation.
Instead, the runtime is limited by the next longest ping pong particles, and the overall runtime is much shorter.
Second, we see a significant amount of overhead distributed among all the ranks in the first part of the runtime. This is the original ping pong particles in the duplicated blocks, but now that overhead is distributed among all the ranks instead of just two.
Of course, the research challenge is identifying and duplicating the necessary data without the post hoc knowledge used here.

\begin{figure}[htb]
\centering
\includegraphics[width=73mm]{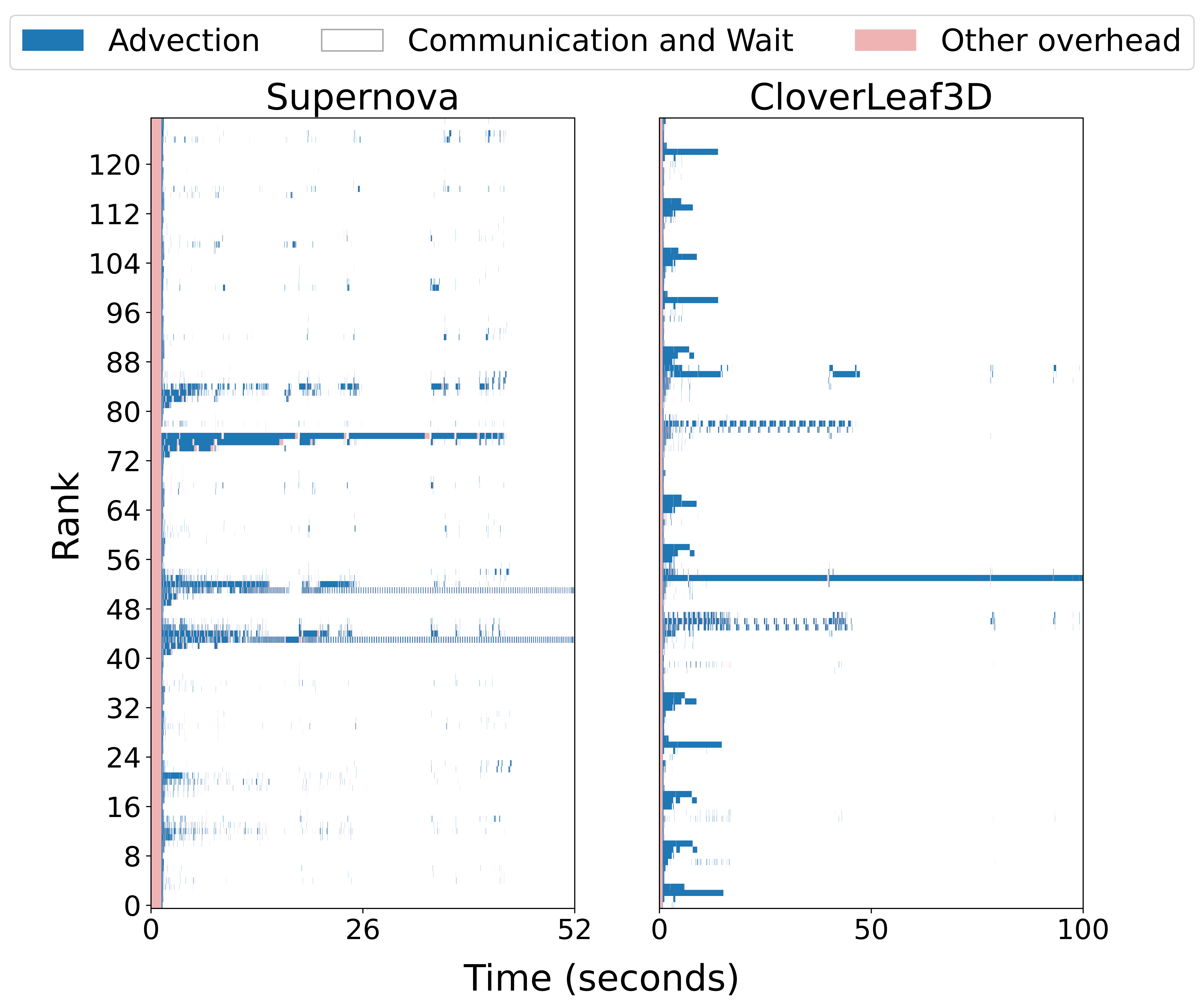}
\caption{Gantt charts of the experiments rerun with the blocks of ping pong particles \emph{merged} together. Supernova (left) has blocks 76 and 77 merged into one large data block and placed on rank 76 whereas Cloverleaf3D (right) has blocks 45 and 53 merged and placed on rank 53.}
\label{fig:discuss:merge}
\vspace{-5pt}
\end{figure}

Another problem we identify with ping pong particles is that a small loop over blocks adds significant overhead during block transition.
Thus, we further hypothesize that we can improve runtime by mitigating or removing this overhead.
One way is to merge blocks or otherwise repartition to remove the ping pong particles.
Figure~\ref{fig:discuss:merge} demonstrates results by merging the problematic blocks.
In this case, when particles enter a rank containing the merged block, they stay there until termination.
Doing so removes the ping pong particles.
What is particularly interesting in this case is the observation that the rank holding the merged block is now single-handedly advecting the majority of the path for these particles.
For the CloverLeaf3D, this means that rank 53 is advecting the majority of the path for 100,000 particles.
And although we can see in Figure~\ref{fig:discuss:merge} that rank 53 is still the bottleneck, the overall runtime is reduced by almost a factor of 2.
One would think this would lead to less balance and longer runtimes, but it instead reduces the overall runtime by removing overhead.
Once again, the research challenge is identifying and repartitioning the data without post hoc knowledge.

\begin{figure}[htb]
\centering
\includegraphics[width=73mm]{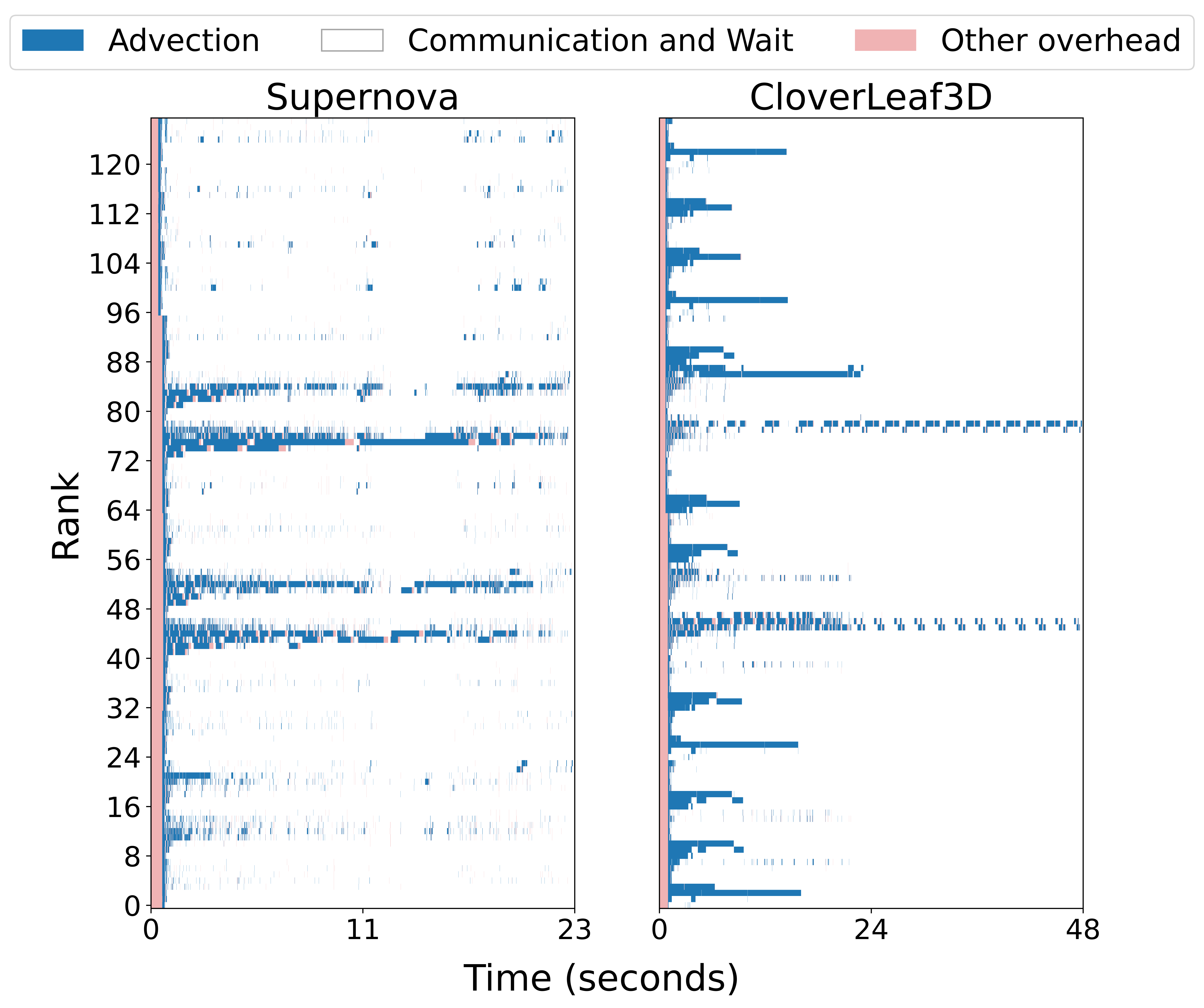}
\caption{Gantt chart of Supernova and CloverLeaf3D dataset with the strategy of early termination for ping pong particles.} 
\label{fig:discuss:earlyterm}
\vspace{-5pt}
\end{figure}

Another solution is to detect particles that are trapped in a vortex and terminate them when they travel back along the vortex.
The early termination strategy is adopted by Peterka et al.~\cite{Peterka2011:IPDPS}.
Their work only terminates a particle that is in the same block at the end of the round.
We need to record the particle path and analyze associated coordinates to detect if they are ping pong particles.
Figure~\ref{fig:discuss:earlyterm} shows results of early termination for Supernova and CloverLeaf3D as a proof of concept. We terminate particles that traverse between two adjacent blocks for two iterations, such as $id1, id2, id1, id2, id1, id2$. After using the early termination strategy, the speed of particle execution is around $4.3\times$ faster for the Supernova dataset and $4.9\times$ faster for the CloverLeaf3D dataset than our original experiments shown in Figure~\ref{fig:results:gantt}.

Although this simple strategy is effective, we should recognize that the result of the particle advection is modified by cutting short many of the trajectories, which could have consequences on the output.
The approach assumes that iterations on a looped trajectory do not need to be repeated, but this is not true for all use cases of particle advection.
Also, just because a particle loops through blocks does not necessarily mean the trajectory is repeating; the trajectory may be a spiral that eventually leads to a different direction.
Besides, detecting looping trajectories requires extra metadata, which adds its own overhead that needs to be evaluated in future work.

\section{Conclusions}
\label{sec:con_future}

Particle advection is a foundational visualization algorithms and core to the analysis of flows that are present in many scientific simulations.
The difficulty in efficient parallelism is well-known and an active area of research.
In this paper, we have presented new results that shed light on the underlying causes of poor performance and provide guidance for further research in addressing these issues.
Our results are derived from a large set of runs on the Summit supercomputer. We varied the runs over three different axes: level of parallelism; data set to capture a wide range of flow characteristics; and number of advection steps to capture a range of particle advection analysis tasks.

Our analysis of these results is derived from analysis from the perspective of the set of MPI processes as well as from the perspective of particles. We provide a model for the execution time in \S~\ref{sec:model} and use it as a basis for analysis of algorithm performance.
We introduced two metrics that quantify the workload imbalance as it evolves throughout algorithm execution.
These metrics highlight the poor efficiency of the POD algorithm, even for ideal use cases (i.e., the Tokamak and the Synthetic datasets).
From the particle-centric analysis, we have shown that the overheads associated with particle movement (not the communication itself) are the source of decreased performance. Further, the size of these groups of particles that are processed together in batches can also have a tremendous impact on scalability. These effects become more pronounced when particles ping pong between a set of adjacent blocks.

In \S \ref{discus:potential_solutions}, we discuss directions for further research to address the performance issues associated with long-running particles and provide some initial evidence of the fruitfulness of these approaches. One approach is block duplication so that the advection of particles, and more importantly, the overheads associated with particle movement, can be amortized over more processors. Another approach is to address the ping pong effect by either termination of such particles (albeit at the risk of introducing errors to flow analysis) or identifying blocks where this occurs and merging them together -- effectively removing the overheads associated with particle movement.

In the future, we plan to (1) extend our preliminary findings in \S \ref{discus:potential_solutions}, (2) refine our models for GPU-enabled particle advection algorithms, and (3) continue refining the approaches for solving the ping pong effect with unsteady flow.

\section{Acknowledgements}
This research used resources of the Oak Ridge Leadership Computing Facility, which is a DOE Office of Science User Facility supported under Contract DE-AC05-00OR22725.
This work was supported in part by the U.S. Department of Energy (DOE) RAPIDS SciDAC project under contract number DE-AC05-00OR22725 and by the Exascale Computing Project (17-SC-20-SC), a collaborative effort of the U.S. Department of Energy Office of Science and the National Nuclear Security Administration.







\bibliographystyle{unsrt}
\clearpage
\bibliography{main.bbl}

\end{document}